\documentclass[letterpaper,12pt]{article} 
\usepackage[]{algorithm2e}
\usepackage{amsmath}
\allowdisplaybreaks
\usepackage[usenames,dvipsnames]{color}
\usepackage[hmargin=0.7in,vmargin=1in]{geometry}
\usepackage[comma,longnamesfirst]{natbib}
\usepackage{booktabs}
\usepackage{amsfonts}
\usepackage{float}
\usepackage{amsmath, amssymb}
\usepackage{graphicx}
\usepackage{epstopdf}
\usepackage{caption}
\usepackage{subcaption}
\usepackage{blindtext}
\usepackage[utf8]{inputenc}
\usepackage{lipsum}
\usepackage{setspace}
\usepackage{pdflscape}
\usepackage{numprint} 
\newtheorem{theorem}{Theorem}

\usepackage{fontenc}
\usepackage{epsfig}
\usepackage{rotating}
\usepackage{hyperref}
\usepackage{multirow}
\usepackage{bm}

\epsfverbosetrue
\usepackage{enumitem}
\newlist{props}{enumerate}{1}
\setlist[props,1]{
	label={\arabic*.},
	leftmargin=*,
	align=left,
	labelsep=5pt,
}

\makeatletter
\def\@seccntformat#1{\@ifundefined{#1@cntformat}%
   {\csname the#1\endcsname\quad}  
   {\csname #1@cntformat\endcsname}
}
\let\oldappendix\appendix 
\renewcommand\appendix{%
    \oldappendix
    \newcommand{\section@cntformat}{\appendixname~\thesection\quad}
}
\makeatother

\usepackage{listings}
\usepackage{color} 
\definecolor{mygreen}{RGB}{28,172,0} 
\definecolor{mylilas}{RGB}{170,55,241}
\usepackage[mathlines,displaymath]{lineno}
\begin{document}
\pagestyle{empty}
\begin{titlepage}
\title{{\sf Variational Bayes Estimation of Discrete-Margined Copula Models with Application to  Time Series}}
\author{Rubén Loaiza-Maya and Michael Stanley Smith}
\date{First Version December 2017\\This Version \today}
\maketitle
\noindent {\small Rub\'{e}n Loaiza-Maya is a PhD student 
and Michael Smith is Chair of Management (Econometrics), both
at Melbourne Business School, University of Melbourne.
Correspondence should be directed to Michael Smith at
{\tt mike.smith@mbs.edu}. }

\newpage
\begin{center}
{\LARGE {\sf  Variational Bayes Estimation of Discrete-Margined Copula Models with Application to  Time Series}}\\
\vspace{15pt}
{\bf Abstract}
\end{center}
\vspace{-10pt}
\noindent
We propose a new variational Bayes estimator for high-dimensional
copulas with discrete, or a combination of discrete and continuous, margins. The method
is based on a variational approximation to a tractable augmented posterior, and is faster than previous likelihood-based approaches.
We use it to estimate drawable vine copulas for univariate and
multivariate Markov ordinal and mixed time series.
These have dimension $rT$, where $T$ is the number of observations
and $r$ is the number of series,
and are difficult to estimate using previous methods. 
The vine pair-copulas are carefully selected to allow for heteroskedasticity, 
which is a feature of most
ordinal time series data. When combined with flexible margins, the resulting 
time series models also allow for other common features of ordinal data, such as
zero inflation, multiple modes and under- or over-dispersion.
Using six example series, we illustrate both the flexibility of the time series copula models, and  
the efficacy of the variational Bayes estimator for copulas of up to 792 dimensions 
and 60 parameters. 
This far exceeds the size and complexity of copula models for 
discrete data that can be estimated
using previous methods. 
\vspace{3cm}


\end{titlepage}

\newpage
\pagestyle{plain}
\newpage
\setlength{\abovedisplayskip}{0.15cm}
\setlength{\belowdisplayskip}{0.15cm}
\vspace{-15pt}
\section{Introduction}
\vspace{-10pt}
Copula models for $m$ discrete-valued variables are difficult to estimate because their 
likelihood involves $2^m$ evaluations of the copula function, so that it is computationally intractable for even moderate dimensions. To avoid this problem, 
\cite{pitt2006} and \cite{SmithKhaled2012} propose using Markov chain Monte Carlo (MCMC)  data augmentation, where a tractable 
augmented likelihood is employed instead. 
However, this approach becomes slow
for copulas with higher dimensions and sample sizes.
\cite{gunawan2016computationally} were the first to
suggest using a variational Bayes (VB) estimator as an alternative.
Their method --- based on that in~\cite{tran2017variational} and labeled VBIL --- uses an unbiased estimate to the intractable likelihood computed
using importance sampling. However, as the copula dimension
or the number of copula parameters increase, computing the unbiased estimate of the likelihood
via importance sampling
also makes this method computationally infeasible. 

In this paper, we propose a new
VB estimator for copulas with a substantially
higher dimension and number of parameters
than can be estimated by either MCMC data augmentation or VBIL. It
uses a variational approximation to the tractable augmented likelihood of~\cite{SmithKhaled2012},
instead of the intractable likelihood. We label our method
VBDA for `variational Bayes data augmentation', and propose several
variational approximations which  balance computational efficiency against accuracy. 
While the new VBDA method is a general
approach to estimate large copula models with one or more discrete margins, 
we employ it here to estimate 
multivariate times series copula models for ordinal, or a combination
of continuous and ordinal (ie. `mixed'), time series variables. The models combine
arbitrary
time-invariant margins with a copula that
captures serial and cross-sectional dependence jointly~\citep{beare15,Smith2015}. These
copulas are  
challenging to estimate because they have dimension $rT$, where $T$ is the number of 
time series observations
and $r$ is the number of series. 
We first show that VBDA is accurate when compared to the
exact posterior computed using
(much slower) MCMC data augmentation for some
univariate ($r=1$) ordinal time series. 
We then employ it to estimate multivariate times series copulas,
where MCMC cannot be used in practice.

Ordinal time series data arise in many fields,
such as criminology~\citep{mohler2013modeling}, marketing~\citep{Ravishanker2016} and finance \citep{bien2011inflated,aktekin2013assessment}. 
These series often exhibit over- or under-dispersion, multiple modes, truncation and zero-inflation
in the margin, along with serial correlation in the level and also conditional variance (ie. heteroskedasticity). There is an extensive literature
on models that can capture one or more of these
features in univariate series;  see~\cite{harvey1989time} and \cite{davis2016handbook} for some examples. 
In contrast, extension to multiple series is difficult and less common; for examples,  see~\cite{heinen2007multivariate,pedeli2011bivariate} and \cite{aktekin2017}. 
In comparison, time series copula models
allow for over- or under-dispersion, multiple modes, truncation and zero inflation in a time 
series through the choice
of an arbitrary margin. They also extend readily to multiple ordinal or mixed time series. 
However, a
major challenge is selecting a high-dimensional copula that 
can capture both persistence in the mean and  heteroskedasticity parsimoniously. To do so, we employ a drawable vine (D-vine) \citep{Aas2009182},
which~\cite{beare15} and \cite{Smith2015} show is parsimonious for Markov and stationary multivariate time series.
Following~\cite{LoaizaSmithManee2017}, the component
`pair-copulas' are carefully selected to capture heteroskedasticity in the series.
This is important because it is a feature exhibited by most ordinal time series.

\cite{heinen2007multivariate} use low-dimensional copulas 
to capture cross-sectional dependence between multiple ordinal-valued time series. 
However, this is different from what we propose here, 
where we employ intrinsically high-dimensional copulas to capture both serial and cross-sectional
dependence jointly.  In early work, \cite[Sec. 8.2]{joe1997} outlined a copula-based Markov time
series model for ordinal data, which \cite{nikomentz17} extend to multivariate panel data using 
a low-dimensional elliptical copula to capture contemporaneous cross-sectional dependence. While this model is
parsimonious and tractable, it does not allow for direct dependence
between lagged values of series. This is often
important in the time series modeling of multivariate continuous data, and we find it is also 
important in our multivariate examples here.


The efficacy of both the VBDA estimator and the proposed copula time series model 
is illustrated using six example time series. The first three are monthly counts of  
murder, attempted murder and manslaughter in the Australian state of New South Wales. 
Univariate time series copula models
show that both counts of murder and attempted murder exhibit
serial dependence, including heteroskedasticity. When compared to the (effectively exact)
posterior computed using MCMC data augmentation, the VBDA estimates prove highly 
accurate, yet are much faster to compute. The fourth example is a 
binary-valued time series 
simulated from an auto-logistic regression with very high serial dependence.
This is an extreme test,
for which MCMC data augmentation fails, yet VBDA gives good results.   
The VBDA estimator is then applied to a trivariate 
time series copula model of the three crime count series.
The 792-dimensional copula  
captures a rich multivariate serial dependence
structure, and is difficult to estimate using MCMC data augmentation in reasonable time.
The last example illustrates the mixed margin case, where 
the bivariate serial dependence structure of monthly counts of U.S. bankruptcies and 
the VIX
(which is a continuous-valued index of stock market volatility) is estimated. The marginal distribution 
of the VIX is highly irregular, making the copula model attractive because it can be
modeled nonparametrically. The estimated 658-dimensional copula  
captures heteroskedasticity in both series, and indicates that the VIX is a leading
indicator of U.S. bankruptcies. These copula models cannot be estimated using the 
VBIL method as outlined in~\cite{gunawan2016computationally} in reasonable time.

The paper outline is as follows. Section \ref{section:2} outlines copula models for discrete
data, and the D-vine copula for univariate time series. Section~\ref{section:3} presents
our new VBDA method, including different variational approximations to the augmented posterior,  MCMC data augmentation, and the four univariate examples.
Section \ref{section:4} extends the vine copula model to the case of multiple ordinal time series,
and Section \ref{section:5} to a mixture of ordinal and continuous series. Section~\ref{section:6} concludes.
\vspace{-15pt}
\section{Copula Model}\label{section:2}
\vspace{-10pt}
\subsection{Copula with Discrete Margins}\label{section:2.1}
\vspace{-5pt}
Following \cite{Sklar1959}, the joint distribution function of a discrete-valued random vector
$\bm{Y}=(Y_1,\ldots,Y_T)$ can be written as
 \begin{equation}\label{Eq:C(u)}
F(\bm y|\bm\theta) = C(\bm u|\bm\theta),
 \end{equation}
where $\bm y = \left(y_1,\dots,y_T\right)'$, $\bm u =\left(u_1,\dots,u_T\right)'$, $u_t = G_t\left(y_t\right)$, $G_t$ is the marginal distribution function of $Y_t$, and $C$ is a $T$-dimensional copula function that captures all dependence in $\bm{Y}$. In the copula modeling literature it is usual 
to select a parametric copula for $C$, with parameter vector $\bm{\theta}$.
Because $Y_t\in S$ for a finite or countably infinite set $S$, then $F$ at Equation~(\ref{Eq:C(u)}) is only uniquely defined
on its sample space~\citep{Genest2007}.
Nevertheless, $F$
remains well-defined for any given parametric copula function $C$. Let $b_t=G_t\left(y_t\right)$, and $a_t = G_t(y_t^{-})$ be the left-hand limit of $G_t$ at $y_t$, then the corresponding probability mass function is 
\begin{equation}\label{Eq:Likeli}
f(\bm y|\bm\theta) = \Delta_{a_1}^{b_1}\dots\Delta_{a_T}^{b_T}C\left(\bm \omega|\bm\theta\right)\,,
\end{equation}
where the difference notation of \cite[p.~43]{Nelsen2006} is employed with vector of
differencing variables $\bm{\omega}$. 
Direct computation of Equation~(\ref{Eq:Likeli}) is impractical in higher dimensions because
it involves evaluation of $C$ a total of $2^T$ times. 
However, following \cite{SmithKhaled2012}, likelihood-based estimation can be undertaken by
introducing a latent vector $\bm{U}=(U_1,\ldots,U_T)'$, such that $(\bm{Y}',\bm{U})'$ have
augmented density
\begin{equation}\label{Eq:augmentedLik}
f(\bm{y},\bm{u}|\bm{\theta}) = c(\bm{u}|\bm{\theta})\prod_{t=1}^{T}{\cal I}\left(a_t\le u_t< b_t\right)\,,
\end{equation}
with copula density $c\left(\bm u|\bm\theta\right) = \frac{\partial^T}{\partial u_1,\dots,\partial u_T}C(\bm u|\bm\theta)$, and 
the indicator variable ${\cal I}(X)=1$ if $X$ is true,
and ${\cal I}(X)=0$ otherwise. (Note that an alternative notation to the indicator function
here is $\delta_{y_t}(G_t^{-}(u_t))$, 
which is a Dirac mass at $G_t^{-}(u_t)$, with $G_t^{-}$ the 
quantile function of $Y_t$.)
The margin in $\bm{y}$ of Equation~(\ref{Eq:augmentedLik}) 
is the required mass function at Equation~(\ref{Eq:Likeli}). 

When there are multiple independent observations on $\bm{Y}$, as with
the cross-sectional and longitudinal datasets considered in~\cite{SmithKhaled2012}, 
then the augmented likelihood is the product of Equation~(\ref{Eq:augmentedLik}) over the observations. For
the time series case that is the focus of this
paper, the augmented likelihood is given directly by Equation~(\ref{Eq:augmentedLik}).

Last, we note that throughout the paper we denote copula densities with a `$c$', and
density/mass functions that are posteriors 
with `$p$', variational approximations with `$q$', and all others with `$f$'.

\vspace{-10pt}
\subsection{Time Series Copula} \label{sec:dts}
\vspace{-5pt}
We consider the case where 
$\{Y_t\}$ is a strongly stationarity ordinal-valued stochastic process with Markov order $p$.
Then
$G_t$ is time invariant 
and can be written as $G$, and
the 
main challenge in using the copula model at Equation~(\ref{Eq:C(u)}) is the 
selection of $C$ to capture the serial dependence in the 
series. We note that ordinal time series usually
exhibit persistence in both the mean and variance, so that $C$ should capture this feature. 
To do so, we adopt
a D-vine copula~\citep{Aas2009182} with pair-copula components
carefully selected to capture persistence in the first two moments.

In general, a D-vine copula density is equal to
the product of $T(T-1)/2$ bivariate copula densities called pair-copulas. 
However, when the series has Markov order $p$, the number of pair-copulas is
much smaller. Moreover, when the series is also stationary, the number of 
unique pair-copulas is equal to the Markov order $p$~\citep{beare15,Smith2015}.
For $s<t$, by denoting $u_{t|s} = F(u_t|u_{s},\dots,u_{t-1})$,
$u_{s|t} = F(u_s|u_{s+1},\dots,u_{t})$
and $u_{t|t} = u_t$,
this parsimonious D-vine copula density is
\begin{eqnarray}
c^{DV}(\boldsymbol{u}|\bm{\theta}) &= &\prod_{t=2}^T f(u_t|u_{\max(1,t-p)},\ldots,u_{t-1}) \nonumber \\
&= &\prod_{t=2}^{T}\prod_{k = 1}^{\text{min}(t-1,p)}
c_{k+1}\left(u_{t-k|t-1},u_{t|t-k+1};\bm{\theta}_{k+1}\right)\,,
\label{eq:DvineLikeli}
\end{eqnarray}
where $\bm{\theta} = \{\bm\theta_2,\dots,\bm{\theta}_{p+1}\}$ and $c_2,\ldots,c_{p+1}$
are the pair-copula densities. Given $\bm{u}$,
 the arguments $\{u_{t|s},u_{s|t}\,; t=2,\ldots,T\,, s<t\}$ 
are computed using the recursive Algorithm~1 in~\cite{Smith2015}.

\cite{LoaizaSmithManee2017} show that $c^{DV}$ is able to capture persistence
in the variance if one or more $c_k$ allows for concentration of the probability mass in the four quadrants of the unit square. To do so they suggest
 the following mixture of rotated copulas:
\begin{equation}
c^{MIX}(u,v;\bm{\gamma}) = wc^a(u,v;\bm{\gamma}^a)+(1-w)c^b(1-u,v;\bm{\gamma}^b)
\,,\;0\leq w \leq1\,.
\label{eq:cmix}
\end{equation}
Here, $\bm{\gamma}=\{\bm{\gamma}^a,\bm{\gamma}^b,w\}$, 
$0\leq w \leq 1$ is a weight, and $c^a,c^b$ are two parametric bivariate copula densities with non-negative Kendall's tau and parameters $\bm{\gamma}^a$ and $\bm{\gamma}^b$ respectively. 
In our empirical work, for the mixture components $c^a$ and $c^b$ we employ the 
`convex Gumbel' defined as follows.
Let $c^G(u,v;\tau)$ be the density of a Gumbel copula parameterized (uniquely)
in terms of its Kendall tau value $0\leq \tau < 0.99$. (Note that we bound
$\tau$ away from 1 to enhance numerical stability of the D-vine copula.) 
Then the convex Gumbel has a density $c^{cG}$ equal to the convex
combination of that of the Gumbel and its rotation 180 degrees (ie. the
survival copula), so that
\[
c^{cG}(u,v;\tau,\delta)=\delta c^G(u,v;\tau) + (1-\delta)c^G(1-u,1-v;\tau)\,,
\]
with $0\leq \delta \leq 1$.
When employed for $c^a$ and $c^b$ in Equation~(\ref{eq:cmix}), it gives a
five parameter bivariate copula with $\bm{\gamma}^a=(\delta^a,\tau^a)$, 
$\bm{\gamma}^b=(\delta^b,\tau^b)$,
and a density $c^{MIX}$ that is equal 
to a mixture of all four 90 degree rotations of the Gumbel copula. 
We use independent uniform priors on the elements of $\bm{\gamma}$ 
in our empirical work. 

To measure 
the level of serial dependence captured by our copula model, we use the Spearman's correlation
between $Y_s$ and $Y_t$ for $s<t$. Following
\cite{Genest2007}, for ordinal-valued variables this is
\begin{align*}
\rho_k & =3\sum_{y_s\in S}\sum_{y_t\in S}g\left(y_s\right)g\left(y_t\right) \left(\bar C_{k}\left(b_s,b_t\right) +\bar C_{k}\left(b_s,a_t\right)  + \bar C_{k}\left(a_s,b_t\right) +\bar C_{k}\left(a_s,a_t\right)\right) -3\,,
\end{align*}
where $\bar C_{k}$ is the copula function of the distribution of 
$(Y_s,Y_t)$, which only varies with $k=t-s$ when ${Y_t}$ is stationary~\citep{Smith2015}. 
This copula $\bar C_k$ is constructed by 
simulating (many) draws of $\bm{u}$ from $c^{DV}$ using Algorithm~2 in~\cite{Smith2015}, and then constructing the bivariate empirical
copula from the draws of elements $(u_s,u_{s+k})$.

\vspace{-15pt}
\section{Bayesian Estimation}\label{section:3}
\vspace{-10pt}
From Equation~(\ref{Eq:augmentedLik}), the augmented posterior density is
\begin{equation}\label{Eq:postaugmented}
p(\bm{\theta},\bm{u}|\bm{y}) = \frac{f(\bm{y},\bm{u},\bm{\theta})}{f(\bm{y})}=\left( c(\bm{u}|\bm{\theta})p\left(\bm\theta\right)\prod_{t=1}^{T}
{\cal I}\left(a_t\le u_t< b_t\right) \right)/f(\bm{y})\,,
\end{equation}
where $p(\bm{\theta})$ is the prior and $f(\bm{y})$ is the
 marginal likelihood. The augmented posterior above admits $p(\bm{\theta}|\bm y)$ as one of its margins, and is 
tractable (up to proportionality). \cite{SmithKhaled2012} propose a MCMC data augmentation method for its (effectively exact) evaluation. However, 
this MCMC scheme is generally slow, and computationally infeasible for 
high-dimensional copulas.
Variational Bayes (VB) is an alternative inferential method to MCMC, with both methods typically applicable to the same problems.
Here, we use the augmented posterior above to develop a new
VB estimator for $p(\bm{\theta}|\bm y)$. 

\vspace{-10pt}
\subsection{Variational Bayes Estimator}\label{VB}
\vspace{-5pt}
VB makes possible the estimation of copula models with discrete margins, even for copulas in high dimensions and with a large number of parameters.
Here, $p(\bm{\theta},\bm{u}|\bm{y})$ is approximated by a tractable density  $q_{\lambda}\left(\bm{\theta},\bm{u}\right)$ with parameters $\bm{\lambda}$, called the variational
approximation. 
Estimation consists of finding values of $\bm\lambda$ that minimize the Kullback-Leibler divergence  
\[
\text{KL}\left(q_{\lambda}\left(\bm{\theta},\bm{u}\right)||p(\bm{\theta},\bm{u}|\bm{y})\right) =\int \text{log}\left(\frac{q_{\lambda}\left(\bm{\theta},\bm{u}\right)}{p(\bm{\theta},\bm{u}|\bm{y})}\right)q_{\lambda}\left(\bm{\theta},\bm{u}\right)\mbox{d}\bm{\theta}\mbox{d}\bm{u}\,.
\]
This can be shown~\citep{jordan1999introduction,ormerod2010explaining} to correspond to maximizing the lower bound of the logarithm of the marginal likelihood $\text{log} \ p(\bm y)$, given by
\[
\mathcal{L}\left(\bm{\lambda}\right)=\int \text{log}\left(\frac{p(\bm{\theta})
	f(\bm{y},\bm{u}|\bm{\theta})}{q_{\lambda}(\bm{\theta},\bm{u})}\right)q_{\lambda}\left(\bm{\theta},\bm{u}\right)\mbox{d}\bm{\theta}\mbox{d}\bm{u}\,.
\]
In selecting $q_\lambda$, it is common to assume independence between some or all
parameters  \citep{mcgrory2007variational,wand2011mean}, and we do so here 
between $\bm \theta$ and $\bm{U}$. The variational approximation we use has density
\begin{equation}
q_{\lambda}\left(\bm{\theta},\bm{u}\right)=
q_{\lambda^a}(\bm{\theta}) q_{\lambda^b}(\bm{u})
\,,\label{eq:vapprox}
\end{equation}
where the density
$q_{\lambda^a}$ has parameters $\bm{\lambda}^a$,
the density $q_{\lambda^b}$ has support on $[a_1,b_1)\times \ldots \times [a_T,b_T)$ and parameters $\bm{\lambda}^b$, and $\bm{\lambda}=\{\bm{\lambda}^a,\bm{\lambda}^b\}$.
The key to the success of our method is the specification of 
$q_{\lambda^a}$ and $q_{\lambda^b}$, which we discuss in detail later.

We follow \cite{paisley2012,nott2012,hoffman2013,ranganath2014black} and
others and use
stochastic gradient ascent (SGA) methods to maximize
 $\mathcal{L}\left(\bm{\lambda}\right)$.
 This approach only requires that (i)~generation from $q_{\lambda}(\bm{\theta},\bm u)$ 
 is possible, and that (ii)~the target distribution is tractable and can be evaluated up to proportionality. 
 Condition~(i) is met by our choices for  $q_{\lambda^a}$ and $q_{\lambda^b}$ outlined below.
 Condition~(ii) is met because the augmented posterior is 
 tractable, whereas $p(\bm{\theta}|\bm{y})$ based on Equation~(\ref{Eq:Likeli}) is
 not. To implement SGA, initial values for the parameters, $\bm \lambda^{(0)}$, are selected and then the lower bound is sequentially optimized by
 values $\bm\lambda^{(1)},\bm{\lambda}^{(2)},\ldots$ obtained by the updating formula
$$\bm\lambda^{(k+1)} = \bm\lambda^{(k)}+\rho^{(k)}\widehat{\nabla_{\lambda}\mathcal{L}\left(\bm{\lambda}^{(k)}\right)}\,.$$
Here, $\widehat{\nabla_{\lambda}\mathcal{L}\left(\bm{\lambda}^{(k)}\right)}$ is an unbiased estimate of the lower bound's gradient 
 $\nabla_{\lambda}\mathcal{L}\left(\bm{\lambda}^{(k)}\right)$, and $\rho^{(k)}$ is the learning rate, set using the ADADELTA method described in the Appendix \ref{Append:Adadelta}.
To compute the gradient, SGA methods resort to the ``log-derivative trick'' ($E_q\left(\nabla_{\lambda}\text{log}\ q_{\lambda}\left(\bm{\theta},\bm u\right)\right) = 0$), and show that 
\begin{equation}\label{Eq:LogTrick}
\nabla_{\lambda}\mathcal{L}\left(\lambda\right)=E_q\left( \nabla_{\lambda}\text{log}\ q_{\lambda}\left(\bm{\theta},\bm{u}\right)\{\text{log}\ h(\bm{\theta,\bm{u}})-\text{log}\ q_{\lambda}\left(\bm{\theta},\bm{u}\right)\}\right)\,,
\end{equation}
with $p(\bm{\theta},\bm{u}|\bm{y})\propto f(\bm{y},\bm{u}|\bm{\theta})
p(\bm{\theta})=h(\bm{\theta},\bm{u})$, and $E_q$ is the expectation with respect to $q_{\lambda}\left(\bm{\theta},\bm{u}\right)$. Notice from Equation (\ref{Eq:LogTrick}) that an unbiased estimate is  
$\widehat{\nabla_{\lambda}\mathcal{L}\left(\bm{\lambda}\right)}= \left(g_{\lambda_1},\dots,g_{\lambda_m}\right)'$, where
$$g_{\lambda_i}=\frac{1}{S}\sum_{s=1}^{S}\left(\text{log}\ h\left(\bm\theta_s,\bm{u}_s\right)-\text{log}\ q_{\lambda}\left(\bm{\theta}_s,\bm{u}_s\right)\right)\nabla_{\lambda_i}\text{log}\ q_{\lambda}\left(\bm{\theta}_s,\bm{u}_s\right)\,,$$
with $m$ as the number of elements in $\bm{\lambda}$. 
An advantage of our choice of variational approximation is that 
the $i$th element of the gradient $\nabla_{\lambda}\text{log}\ q_{\lambda}\left(\bm{\theta},\bm{u}\right)$
simplifies to
$\nabla_{\lambda_i}\text{log}\ q_{\lambda}\left(\bm{\theta},\bm{u}\right)= 
\nabla_{\lambda_i}\text{log}\ q_{\lambda^a}\left(\bm{\theta}\right)+ 
\nabla_{\lambda_i}\text{log}\ q_{\lambda^b}\left(\bm{u}\right)$.
For $s=1,\ldots,S$, the values 
$\bm\theta_s\sim q_{\lambda^a}\left(\bm{\theta}\right)$ and $\bm{u}_s=(u_{1,s},\ldots,u_{T,s})\sim q_{\lambda^b}(\bm{u})$ 
on $[a_1,b_1)\times \cdots \times [a_T,b_T)$. 

Algorithm \ref{alg:VB} presents how the SGA optimization works within variational Bayes. Step (1b) is based on the work by \cite{tran2017variational}, which employs a vector of control variates, $\bm{\varsigma}$, for variance reduction of the unbiased estimate of the gradient. The stopping rule is commonly set as a fixed number of SGA steps taken \citep{ong2017gaussian}.
\begin{algorithm}[h]
	\hrulefill\\
	Initialize $\bm\lambda^{(0)}$ and set $k=0$. 
	\begin{itemize}
		\item[1.] \begin{itemize}
			\item[(a)] Generate $\left(\bm\theta^{(k)}_s,\bm{u}^{(k)}_s\right)\sim q_{\lambda^{(k)}}\left(\bm{\theta},\bm{u}\right)$ for $s = 1,\dots,S$	    
			\item[(b)] Estimate $\bm{\varsigma}^{(k)} =
			\left(\varsigma_1^{(k)},\dots,\varsigma_m^{(k)}\right)'$ with
			$$ \varsigma_{i}^{(k)}=\frac{\text{Cov}\left(\left[ \log h\left(\bm{\theta},\bm{u}\right)-\log q_\lambda(\bm{\theta},\bm{u}) \right]
				\nabla_{\lambda_i}\log q_{\lambda}\left(\bm{\theta},\bm{u}\right),
				\nabla_{\lambda_i}\log q_{\lambda}(\bm{\theta},\bm{u})\right)}{\text{Var}\left(\nabla_{\lambda_i}\text{log}\ q_{\lambda}\left(\bm\theta,\bm u\right)\right)}$$
			Cov(.) and Var(.) are sample estimates of covariance and variance based on the S samples from step (a).
			\item[(c)] $k = k+1$. 
		\end{itemize}
		\item[2.]Repeat until some stopping rule is satisfied
		\begin{itemize}
			\item[(a)] Generate $\left(\bm\theta^{(k)}_s,\bm{u}^{(k)}_s\right)\sim q_{\lambda^{(k)}}\left(\bm{\theta},\bm{u}\right)$ for $s = 1,\dots,S$	     	   	 
			\item[(b)] Compute $\widehat{\nabla_{\lambda}\mathcal{L}\left(\bm{\lambda}^{(k)}\right)}= \left(g_{\lambda_1}^{(k)},\dots,g_{\lambda_m}^{(k)}\right)'$ with \\
			\ \\
			$g_{\lambda_i}^{(k)}=\frac{1}{S}\sum_{s=1}^{S}\left(\text{log}\ h\left(\bm\theta^{(k)}_s,\bm u^{(k)}_s\right)-\text{log}\ q_{\lambda}\left(\bm{\theta}_s^{(k)},\bm u^{(k)}_s\right)-\varsigma_i^{(k-1)}\right)\nabla_{\lambda_i}\text{log}\ q_{\lambda}\left(\bm{\theta}_s^{(k)},\bm u^{(k)}_s\right)$
			\item[(c)] Estimate $\bm{\varsigma}^{(k)} = \left(\varsigma_1^{(k)},\dots,\varsigma_m^{(k)}\right)'$ as in Step 1(b).
			\item[(d)] Compute $\Delta\bm{\lambda}^{(k)}$ using the ADADELTA method.
			\item[(e)] Set  $\bm\lambda^{(k+1)}=\bm\lambda^{(k)}+\Delta\bm{\lambda}^{(k)}$.
			\item[(f)] $k = k+1$
		\end{itemize} 
	\end{itemize}\hrule
	\caption{Variational Bayes estimation algorithm with control variates and ADADELTA learning rate for an the augmented posterior.}
	\label{alg:VB}
\end{algorithm}

\vspace{-10pt}
\subsection{Variational Approximation}\label{sec:va}
\vspace{-5pt}
Key to developing an effective VB estimator is the selection of $q_{\lambda^a}$ and $q_{\lambda^b}$ in 
Equation~(\ref{eq:vapprox}) that balance tractability and accuracy. 
We first outline
three choices for $q_{\lambda^b}$, after which we then detail that for $q_{\lambda^a}$.

\vspace{-10pt}
\subsubsection{Approximation for $U$}
\vspace{-5pt}
To guide our choice for $q_{\lambda^b}$, we derive the marginal posterior of
$\bm{U}$ in Theorem~\ref{theorem1}.
\begin{theorem}\label{theorem1}
If $(\bm{\theta},\bm{U})$ have the augmented posterior density function at Equation~(\ref{Eq:postaugmented}), then: 
\begin{itemize}
	\item[(a)] The joint density $p(\bm{u}|\bm{y})=\tilde{c}(\bm{u}) \prod_{t=1}^T {\cal I}(a_t\leq u_t < b_t)/f(\bm{y})$,
	where $\tilde c(\bm{u})=\int c(\bm{u}|\bm{\theta})p(\bm{\theta})\mbox{d}\bm{\theta}$ is a copula density and $f(\bm{y})$ is the marginal likelihood; and,
	\item[(b)]
	the marginal density $p(u_t|\bm{y})\propto{\cal I}(a_t\leq u_t < b_t)\int A(u_t|\bm{\theta})p(\bm{\theta}|\bm{y})\mbox{d}\bm{\theta}$, where 
	\[
	A(u_t|\bm{\theta})=\Delta_{a_1}^{b_1}\cdots \Delta_{a_{t-1}}^{b_{t-1}} 
	\Delta_{a_{t+1}}^{b_{t+1}} \cdots \Delta_{a_T}^{b_T} H(v_1,\ldots,v_{t-1},u_t,v_{t+1},\ldots,v_T|\bm{\theta})\,, 
	\]
	$H(\bm{u}|\bm{\theta})=\int c(\bm{u}|\bm{\theta})\mbox{d}\bm{u}_{s\neq t}$  and $\bm{u}_{s\neq t}=(u_1,\ldots,u_{t-1},u_{t+1},\ldots,u_T)$.
\end{itemize}
{\bf Proof}: See Appendix~\ref{append:proof}.
\end{theorem}
We make two observations on the  posterior of $\bm{U}$.
First, if the elements of $\bm{Y}$ are independent, then $\tilde{c}(\bm{u})=1$ is the density 
of an independence copula, and $p(\bm{u}|\bm{y})\propto \prod_{t=1}^T {\cal I}(a_t\leq u_t < b_t)$, so that each element is independent uniform. Second, as $(b_t-a_t) \rightarrow 0$ for 
all $t$, then $p(\bm{u}|\bm{y})\xrightarrow[]{d} \prod_{t=1}^T \delta_{u_t}(a_t)$. That is, as the data $\bm{Y}$ becomes
`closer to continuous', the posterior approaches a degenerate distribution with point mass at  $\bm{u}=(a_1,\ldots,a_T)'$.
 
Armed with these observations, our first choice is simply independent uniforms:
\[
\mbox{VA1:} \;\;\; q_{\lambda^b}(\bm{u})=\prod_{t=1}^T \frac{1}{b_t-a_t}{\cal I}(a_t \leq u_t < b_t)\,, \mbox{ so that } {\bm \lambda}^b=\emptyset\,.
\]
We expect VA1 to be more accurate for data with low dependence (although we find it still 
works well for even quite dependent data).
The next two approximations are based on normal distributions for a transformation 
of $\bm{U}$. Let
$Z_t=\Phi^{-1}\left((U_t-a_t)/(b_t-a_t)\right)$, and
$\bm{Z}=(Z_1,\ldots,Z_T)'\sim N(\bm{\eta},\Omega)$, with
$\bm{\eta}=(\eta_1,\ldots,\eta_T)'$ and $\Phi$ the standard normal distribution
function. The Jacobian of this transformation is 
$J_{\bm{Z}\rightarrow \bm{U}}=\prod_{t=1}^T \left((b_t-a_t)\phi(z_t)\right)^{-1}$, with
$\phi$
the standard normal density and $z_t=\Phi^{-1}\left((u_t-a_t)/(b_t-a_t)\right)$.
Our second choice for $q_{\lambda^b}$ assumes 
$\Omega=\mbox{diag}(\omega^2_1,\ldots,\omega^2_T)$, so that
\[
\mbox{VA2:} \;\;\; q_{\lambda^b}(\bm{u})=
	\prod_{t=1}^T \frac{\phi_1(z_t;\eta_t,\omega^2_t)}{(b_t-a_t)\phi(z_t)}\,,\mbox{ with } 
	\bm{\lambda}^b=\{\bm{\eta},\log \omega_1,\ldots,\log \omega_T\}\,,
\]
and $\phi_1(z_t;\eta_t,\omega^2_t)$ is the density of a $N(\eta_t,\omega^2_t)$ distribution
evaluated at $z_t$. Note that VA2 nests VA1.
We find this an effective mean field approximation that is accurate for a wide range of data,
and very fast to work with. 

For $\bm{Y}$ that exhibits extreme dependence,
our third choice allows for the elements
of $\bm{U}$ to be dependent\footnote{We 
are grateful to an anonymous
referee who suggested that this may be an important consideration.}
by adopting a non-diagonal (but sparse) precision matrix $\Omega^{-1}$.
For time series copulas, we set $\Omega^{-1}=LL'$, with $L$ a band one lower triangular
Cholesky factor. This corresponds to an approximation $q_{\lambda^b}$ with the dependence
structure of a (non-stationary) first order Markov process for $\{U_t\}$. The density
\[
\mbox{VA3:} \;\;\; q_{\lambda^b}(\bm{u})=
 \frac{\phi_T(\bm{z};\bm{\eta},(LL')^{-1})}{\prod_{t=1}^T(b_t-a_t)\phi(z_t)}\,,\mbox{ with } 
\bm{\lambda}^b=\{\bm{\eta},L\}\,,
\]
where only the free elements of $L$ are variational parameters. If the lower triangular first band
of $L$ contains only zeros, then VA3 reduces to VA2. Both approximations are fast to generate
from by first generating $\bm{z}$ from normals, and then transforming to 
$\bm{u}$. The gradients $\nabla_{\lambda^b} \log q_{\lambda^b}(\bm{u})$ required to 
implement Steps~1(b) and~2(b) of Algorithm~\ref{alg:VB} are available in
closed form; see Appendix~\ref{append:nabla}.

Note, as $(b_t-a_t) \rightarrow 0$ for all $t$, all three approximations 
become exact. The accuracy of $q_{\lambda^b}$ is important because
it can also increase the accuracy of the variational approximation of $p(\bm{\theta}|\bm{y})$.

\vspace{-10pt}
\subsubsection{Approximation for $\theta$}
\vspace{-5pt}
Denoting the number of parameters in $\bm{\theta}$ as $n$, the most popular choice for 
$q_{\lambda^a}$ is the density of a $N(\bm{\mu},\Sigma)$ distribution, because it is quick to
generate from and the gradient $\nabla_{\lambda^a}\log q_{\lambda^a}\left(\bm{\theta}\right)$
is available in closed form~(\citealt{opper2009variational,challis2013gaussian,
titsias2014doubly,kucukelbir2016automatic,salimans2013fixed}).
To ensure $\Sigma$ is positive definite, $\bm{\lambda}^a$ is typically a convenient
re-parametrization of $\bm{\mu}$ and $\Sigma$. 
In applications where $\bm{\theta}$ has a large number of elements, a sparse representation of $\Sigma$ helps to improve the accuracy of the gradient estimate and its speed of computation. We follow~\cite{ong2017gaussian}, who suggest the factor representation of the covariance matrix $\Sigma = B'B+D$, where the matrix $B$ is of dimension $n\times K$, $K$ is the number of factors and $K<<n$. All the elements in the upper triangle of $B$ are set to zero. $D$ is a $n\times n$ diagonal matrix such that $D_{i,i} = d_i^2$, where $d_i$ is the $i^{th}$ element of the vector $\bm d$. \cite{ong2017gaussian} derives the gradient for this case, and shows it is
fast to compute; see also Appendix~\ref{append:nabla}. In our empirical work, we
compare the accuracy of the 
approximations for various values of $K$ and find low values adequate.

\vspace{-10pt}
\subsubsection{Discussion of Alternative VB Approximations}
\vspace{-5pt}
\cite{gunawan2016computationally} suggest using an unbiased estimator of the
intractable likelihood in Equation~(\ref{Eq:Likeli}) computed using importance sampling. This involves
drawing $N_{IS}$ values of $\bm{u}$, at which $c(\bm{u}|\bm{\theta})$ is repeatedly evaluated.
Whenever evaluating the copula density is computationally intensive --- such as for $c^{DV}$ here or with 
other high-dimensional or complex copulas ---  this will be many times slower than our approach.

Variational approximations to posteriors augmented with latent variables
have proven successful in a number of other models; see~\cite{tan2017}, \cite{hui2017} and~\cite{ong2017gaussian}
for some recent examples. However, \cite{neville14} and others observe that assuming independence 
between the latent variables in such an approximation may lead to poor inference in some circumstances. This motivates
VA3, although in our
empirical work VA2 proves almost as accurate and several times faster.
Key to using VA3 for other copula models is the adoption of
an appropriate parsimonious matrix $\Omega$, or its inverse. Last, we mention
it is also possible to employ
a Gaussian approximation with factor covariance 
structure for the vector $(\bm{\theta},\bm{Z})$.  This may improve the accuracy of the approximation
for some copulas, but will introduce an extra $KT$ variational parameters
(ie. the extra factor loadings),
slowing estimation down substantially. Our empirical work suggests that our proposed variational approximations strike a balance between computation
time and accuracy.

\vspace{-10pt}
\subsection{Data Augmentation}\label{sect:DA}
\vspace{-5pt}
We now outline MCMC data augmentation, tailored for the parsimonious
D-vine copula in Section~\ref{sec:dts}. Key to implementation is
the evaluation of the conditional
densities and distribution functions below.
If $t_0=\max(t-p,1)$ and 
$t>1$, then
\begin{eqnarray*}
	f(u_t|u_{t_0},\ldots,u_{t-1}) &= &\prod_{k=1}^{\min(t-1,p)}c_{k+1}(u_{t-k|t-1},u_{t|t-k+1};\bm{\theta}_{k+1})\,\mbox{ and}\\
	F(u_t|u_{t_0},\ldots,u_{t-1}) &= &h_{t_0,t}\circ h_{t_0+1,t}\circ \cdots \circ h_{t-1,t}(u_t)\,,
\end{eqnarray*}  
where $h_{s,t}(u)=h_{t-s+1}^1(u|u_{s|t-1})=\frac{\partial}{\partial v}C_{t-s+1}(v,u)\Big|_{v = u_{s|t-1}}$ is the conditional pair-copula function. This is given in
Appendix~C1 of~\cite{LoaizaSmithManee2017} for the mixture copula
defined at Equation~(\ref{eq:cmix}).

The values $\bm{u}$ are integrated out of the augmented posterior
as part of an MCMC sampling scheme. The scheme generates from the conditional
posteriors~(1)~$p(\bm{u}|\bm{y},\bm{\theta})$,
and~(2)~$p(\bm{\theta}|\bm{u})$. 
Given the values $\bm{u}$, step~(2) can be undertaken using (adaptive) random walk Metropolis-Hastings (MH), where
$\bm{\theta}_{k}$ is generated conditional on 
$\{\bm{\theta}\backslash \bm{\theta}_k\}$ for $k=2,\ldots,p+1$.
However, step~(1) is more involved, with the latent variables $\bm{u}$ 
generated jointly using a MH
step. \cite{SmithKhaled2012} suggest using the
proposal density
$\pi(\boldsymbol{u}) = \prod_{t=2}^{T}\pi_t(u_t|u_{t_0},\dots,u_{t-1})\pi_1(u_1)$,
where
$\pi_1(u_1) = \mathcal{I}(a_1\le u_1<b_1)/(b_1-a_1)$
and
\[
\pi_t(u_t|u_{t_0},\dots,u_{t-1}) \propto \frac{f(u_t|u_{t_0},\dots,u_{t-1})\mathcal{I}\left(a_t\le u_t<b_t\right)}
{F(b_t|u_{t_0},\dots,u_{t-1})-F(a_t|u_{t_0},\dots,u_{t-1})}\,.
\]
Therefore,
a proposal iterate
$\bm{u}^{\mbox{\tiny new}}$ can be
obtained from $\pi(\bm{u})$ by generating sequentially from
the univariate densities $\pi_1,\ldots,\pi_T$. Each of these is a
constrained univariate distribution with known distribution function, so that
iterates can be generated easily using the inverse distribution 
method. An advantage of the proposal $\pi$ is that
the
MH acceptance ratio is fast to compute. The probability of
accepting 
$\bm{u}^{\mbox{\tiny new}}=(u_1^{\mbox{\tiny new}},\ldots,u_T^{\mbox{\tiny new}})$ over the previous value 
$\bm{u}^{\mbox{\tiny old}}=(u_1^{\mbox{\tiny old}},\ldots,u_T^{\mbox{\tiny old}})$ is
\[
\min\left(1\,,\, \prod_{t=2}^{T}\frac{F(b_t|u_{t_0}^{\mbox{\tiny new}},\dots,
	u_{t-1}^{\mbox{\tiny new}})-F(a_t|u_{t_0}^{\mbox{\tiny new}},
	\dots,u_{t-1}^{\mbox{\tiny new}})}
{F(b_t|u_{t_0}^{\mbox{\tiny old}},\dots,u_{t-1}^{\mbox{\tiny old}})-
	F(a_t|u_{t_0}^{\mbox{\tiny old}},\dots,u_{t-1}^{\mbox{\tiny old}})}
\right)\,.
\]

In general, this proposal works well. However, for challenging high-dimensional
copulas with highly dependent binary-valued data $\bm{Y}$, we find that the acceptance
rate for this step can be prohibitively low and the MCMC scheme can get stuck. We illustrate
this empirically below. Throughout, we
employ a burnin sample of 10,000 iterates, followed by 
a further 20,000 iterates from which we compute posterior inference.

 \begin{figure}[H]
 	\begin{center}
 		\includegraphics[scale =0.8]{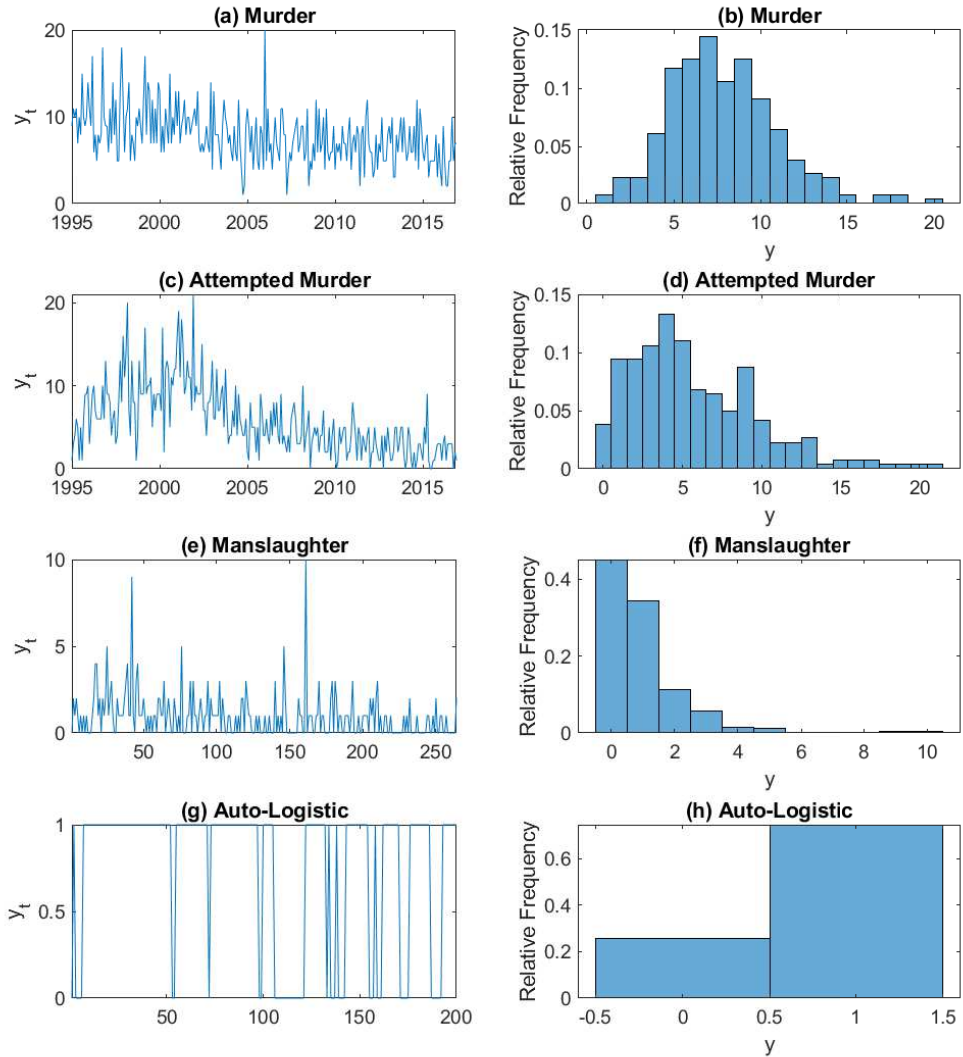}
 	\end{center}
 	\caption{Time series plots and relative frequency histograms of the four univariate ordinal time series. The four rows correspond (from top to bottom) to the Murder, Attempted Murder, Manslaughter and Auto-Logistic examples.}
 	\label{Fig:HomSeries}
 \end{figure}

\vspace{-10pt}
\subsection{Examples}\label{Sect:EmpiricalApUnivariate}
\vspace{-5pt}
We illustrate the copula time series model, and the efficacy of the VB estimator, using four univariate time series examples. The first three are $T=264$ monthly counts of the crimes of Murder, Attempted Murder and Manslaughter
in the Australian state of New South Wales (NSW) between January 1995 and December 2016. The data is sourced from the NSW Bureau of Crime Statistics and Research. The fourth is $T=200$ observations simulated from an 
auto-logistic regression with
$\mbox{Pr}(Y_t=1|y_{t-1})=\mbox{logit}^{-1}(-2.197+4.394y_{t-1})$.
This last example is challenging because (i)~the series is short, (ii)~the data are binary-valued, (iii)~there is very strong serial dependence, and (iv)~there are
many more 1's. 
Figure~\ref{Fig:HomSeries} gives the relative frequency histograms and time series plots of the four series.

\begin{table}
	\centering
	\scalebox{1}{
		\begin{tabular}{lccccccc}
			\toprule
			&       &         $\tau^a>0$          &         $\delta^a$          &         $\tau^b>0$          &         $\delta^b$          &             $w$             &      Kendall's $\tau$       \\ \cline{3-8}
			Murder       & $c_2$ &            0.141            &            0.313            &            0.219            &            0.523            &            0.743            &            0.065            \\
			&       & ({\footnotesize0.04,0.34})  & ({\footnotesize 0.02,0.76}) & ({\footnotesize 0.01,0.81}) & ({\footnotesize 0.06,0.95}) & ({\footnotesize 0.22,0.99}) & ({\footnotesize 0.00,0.13}) \\
			& $c_3$ &            0.161            &            0.610            &            0.190            &            0.516            &            0.782            &            0.097            \\
			&       & ({\footnotesize 0.06,0.34}) & ({\footnotesize 0.14,0.97}) & ({\footnotesize 0.00,0.75}) & ({\footnotesize 0.06,0.95}) & ({\footnotesize 0.27,0.99}) & ({\footnotesize 0.03,0.17}) \\
			& $c_4$ &            0.150            &            0.318            &            0.154            &            0.477            &            0.706            &            0.072            \\
			&       & ({\footnotesize0.04,0.35})  & ({\footnotesize0.02,0.83})  & ({\footnotesize0.00,0.63})  & ({\footnotesize0.04,0.94})  & ({\footnotesize 0.22,0.99}) & ({\footnotesize 0.01,0.14}) \\ \cline{3-8}
			Attempted    & $c_2$ &            0.354            &            0.560            &            0.327            &            0.436            &            0.952            &            0.328            \\
			Murder       &       & ({\footnotesize0.27,0.43})  & ({\footnotesize 0.17,0.91}) & ({\footnotesize 0.01,0.89}) & ({\footnotesize 0.04,0.93}) & ({\footnotesize 0.84,0.99}) & ({\footnotesize 0.24,0.41}) \\
			& $c_3$ &            0.227            &            0.771            &            0.232            &            0.541            &            0.876            &            0.180            \\
			&       & ({\footnotesize 0.15,0.33}) & ({\footnotesize 0.46,0.98}) & ({\footnotesize 0.01,0.78}) & ({\footnotesize 0.07,0.95}) & ({\footnotesize 0.60,0.99}) & ({\footnotesize 0.12,0.25}) \\
			& $c_4$ &            0.155            &            0.676            &            0.225            &            0.512            &            0.810            &            0.099            \\
			&       & ({\footnotesize0.05,0.29})  & ({\footnotesize0.21,0.97})  & ({\footnotesize0.01,0.75})  & ({\footnotesize0.05,0.95})  & ({\footnotesize 0.33,0.99}) & ({\footnotesize 0.03,0.17}) \\ \cline{3-8}
			Manslaughter & $c_2$ &            0.195            &            0.484            &            0.208            &            0.491            &            0.628            &            0.052            \\
			&       & ({\footnotesize0.03,0.50})  & ({\footnotesize 0.04,0.94}) & ({\footnotesize 0.01,0.69}) & ({\footnotesize 0.05,0.94}) & ({\footnotesize 0.12,0.97}) & ({\footnotesize-0.03,0.13}) \\
			& $c_3$ &            0.173            &            0.389            &            0.224            &            0.524            &            0.688            &            0.061            \\
			&       & ({\footnotesize 0.02,0.50}) & ({\footnotesize 0.03,0.90}) & ({\footnotesize 0.01,0.83}) & ({\footnotesize 0.07,0.96}) & ({\footnotesize 0.09,0.98}) & ({\footnotesize-0.01,0.14}) \\
			& $c_4$ &            0.170            &            0.431            &            0.180            &            0.512            &            0.665            &            0.062            \\
			&       & ({\footnotesize0.03,0.42})  & ({\footnotesize0.04,0.91})  & ({\footnotesize0.01,0.66})  & ({\footnotesize0.05,0.94})  & ({\footnotesize 0.14,0.98}) & ({\footnotesize-0.01,0.14}) \\ \bottomrule
		\end{tabular}}
		\caption{Posterior means of the pair-copula parameters for the D-vines fit to the three
			univariate crime count time series, computed using MCMC data augmentation.
			Each pair-copula is a mixture of two convex Gumbels, and the final column
			reports the posterior mean of the overall Kendall's of each pair-copula.}
		\label{tab:ParEstimatesGumUnivariate}
	\end{table}

\begin{figure}[H]
 	\begin{center}
 		\includegraphics[scale =0.8]{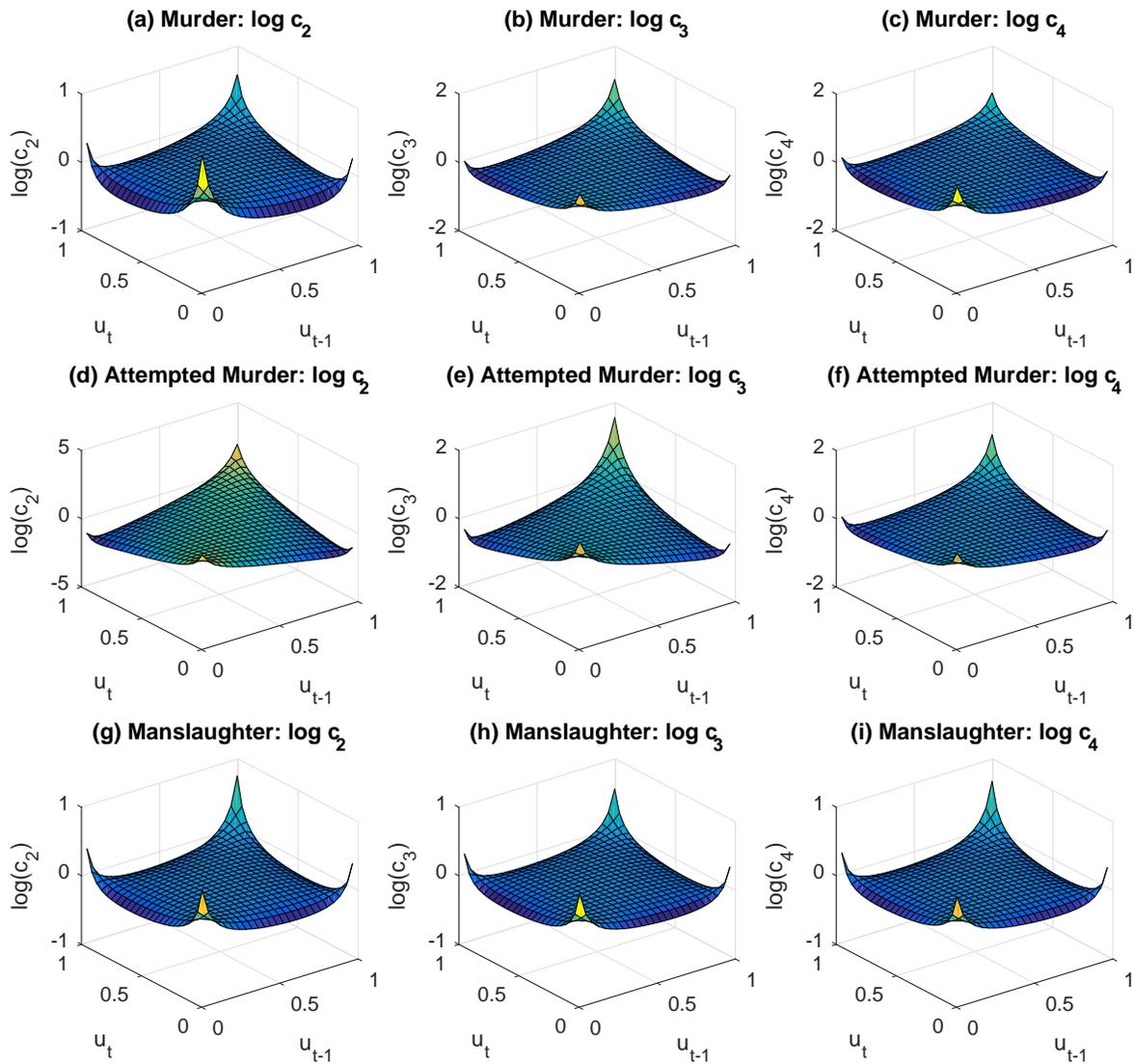}
 	\end{center}
 	\caption{Log-density functions of the pair-copulas at the MCMC posterior mean parameter values for the univariate copula models fit to the three crime series. Columns one to three correspond to the pair-copulas $c_2$, $c_3$ and $c_4$ respectively. Rows one to three correspond to Murder, Attempted Murder and Manslaughter, respectively}
 	\label{Fig:Copulas9}
 \end{figure}

We set $p=3$, and fit the copula using $c^{MIX}$ pair-copula components for $c_2$, $c_3$ and $c_4$. For $c^a$ and $c^b$ we chose convex Gumbels, so that $\bm{\theta}_{k} = \{\tau_{k}^a,\delta_{k}^a,\tau_{k}^b,\delta_{k}^b,w_{k}\}$ for $k = 2,3,4$. This $T$-dimensional
D-vine copula has a total of $n = 15$ parameters.
We set $G$ to the empirical distribution functions in Figure~\ref{Fig:HomSeries}. 
To estimate the copula parameters we first use MCMC data augmentation to compute the exact
posterior as outlined in Section~\ref{sect:DA}.
For the three crime series, between 22\% and 71\% of MH proposals were accepted, but
for the Auto-Logistic example
the sampler became stuck, and estimation failed.
For the crime series,
Table~\ref{tab:ParEstimatesGumUnivariate} reports the posterior means and intervals
 of the copula parameters. To summarize the serial dependence captured by the copula, Table~\ref{tab:spearmansCorrGumbel} reports the posterior of the Spearman correlations
 $\rho_k$ for $k=1,\ldots,3$
 for the three crimes. Correlation is strong for Attempted Murder, but not for Murder and 
 Manslaughter. However, this measures correlation in the level of the series, and not more general dependence, such as in higher order moments. Figure~\ref{Fig:Copulas9} presents the log-densities of the pair-copulas at the posterior mean values. Most of these copula densities are far from uniform, indicating more general serial dependence exists in these series. The pair-copulas have 
 probability mass in the off-diagonal corners of the unit square, which~\cite{LoaizaSmithManee2017}
 show is indicative of serial correlation in conditional variance (ie. heteroskedasticity).

 \begin{table}
 	\centering
 	\begin{tabular}{lccc}
 		\toprule
 		&           $\rho_1$            &           $\rho_2$           &           $\rho_3$           \\ \midrule
 		Murder           &             0.094             &            0.152**             &            0.128**             \\
 		&  ({\footnotesize-0.01,0.19})  & ({\footnotesize 0.05,0.25})  & ({\footnotesize 0.03,0.22})  \\
 		Attempted Murder &             0.458***             &            0.427***             &            0.382***             \\
 		&  ({\footnotesize 0.34,0.56})  & ({\footnotesize 0.33,0.52})  & ({\footnotesize 0.28,0.48})  \\
 		Manslaughter     &             0.057             &            0.073             &            0.076             \\
 		& ({\footnotesize -0.026,0.14}) & ({\footnotesize -0.01,0.16}) & ({\footnotesize -0.00,0.16}) \\ \bottomrule
 	\end{tabular}
 	\caption{The posterior means of the pairwise Spearman correlations $\rho_k$,
 		for $k = 1,2,3$, from the univariate time series copula models fit to the three
 		monthly crime count time series. Correlations with approximate posterior intervals that do not contain zero
 		at the 10\%, 5\% and 1\% level are denoted with `*', `**' and `***', respectively. The values are computed via simulation from the vine copula, as outlined in Section~\ref{sec:dts}.}
 	\label{tab:spearmansCorrGumbel}
 \end{table}

\begin{figure}
 	\begin{center}
 		\includegraphics[scale =1]{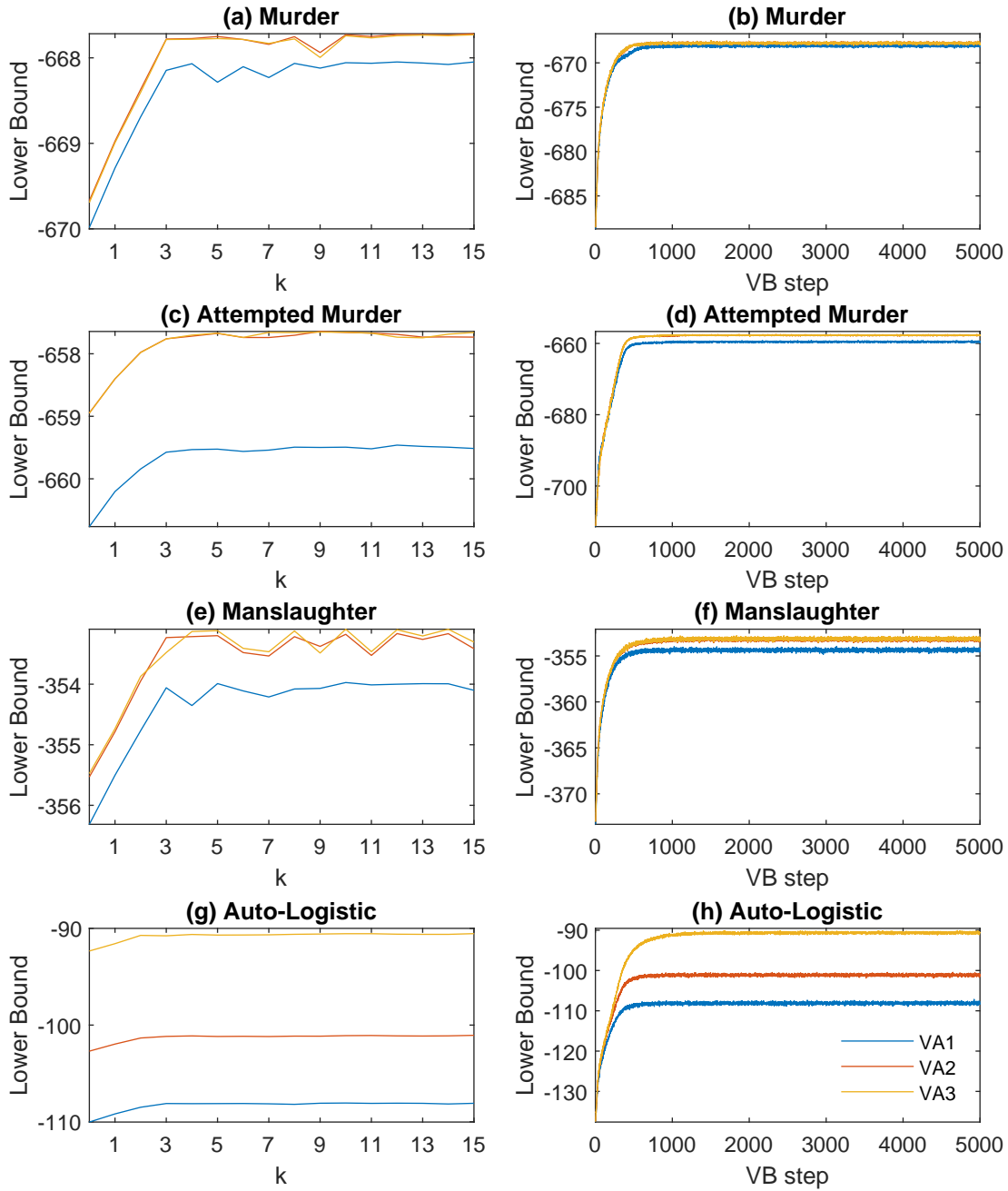}
 	\end{center}
 	\caption{Variational lower bound values ${\cal L}(\bm{\lambda})$ for the VBDA estimators of the univariate time series copula models. From top to bottom, the rows correspond to the Murder, Attempted Murder, Manslaughter and Auto-Logistic examples. 
 		The first column plots  ${\cal L}$ at 
 		the estimate of $\bm{\lambda}$, against the number of factors $K$ 
 		in the factor decomposition  of $\Sigma$. 
 		The second column plots  ${\cal L}$ against VB step for the case of $K=3$ factors. In each panel, 
 		results are given for the
 		VA1 (blue line), VA2 (red line) and VA3 (orange line) approximations.} 
 	\label{Fig:LowerBound}
 \end{figure}

The four D-vines were also estimated using VBDA. 
Because all parameters are bounded between 0 and 1,
we transform them to the real line as $\bm{\theta}_{k} =
 \{\psi(\tau_{k}^a),\psi(\delta_{k}^a),\psi(\tau_{k}^b),\psi(\delta_{k}^b),\psi(w_{k})\}$,
 where $\psi(a) = \text{log}(\frac{a}{1-a})$. Estimation was implemented separately 
for approximations VA1, VA2 and VA3, with $K = 0,\dots,15$ factors. Each
estimator used 5000 SGA steps, and $S=500$ to 
estimate the gradient.
The initial values for $\bm{\lambda}$ are
$\{B=0, D=\sqrt{0.1}I_n, \bm{\mu}=\bm{\mu}_0, \bm{\eta}=\bm{0}, L=I_n\}$, where
$\bm{\mu}_0$ is set to values where the pair-copulas are all independence pair-copulas. 
These initial values are used in all our empirical work, although the results are robust to changes in them.
Figure~\ref{Fig:LowerBound}(a,c,e,g) shows how the lower bound increases with $K$, and any increase is small for $K\geq 3$.
Figure~\ref{Fig:LowerBound}(b,c,f,h) plots the lower bound against SGA step when $K=3$,  suggesting that the SGA algorithm converges within 1000 steps in every case. 
For the three crime series, the lower bounds of 
VA2 and VA3 are almost indistinguishable. For the challenging Auto-Logistic example, 
VA3 --- which allows for dependence in the latent variables --- dominates. However, it is 
difficult to determine how much of the higher lower bound values are attributable 
to an increase in the accuracy of $q_{\lambda^a}$, as opposed to $q_{\lambda^b}$. 
Nevertheless, plots of the pair-copula densities (see Supplementary Materials) for each of
VA1, VA2 and VA3 suggest that VA3 provides a meaningful 
improvement over VA1 and VA2. These also show the VBDA estimates suggest the series
has Markov order one, and high serial dependence; which correspond to the known data generating
process.
 
\begin{figure}
	\begin{center}
		\includegraphics[scale =1]{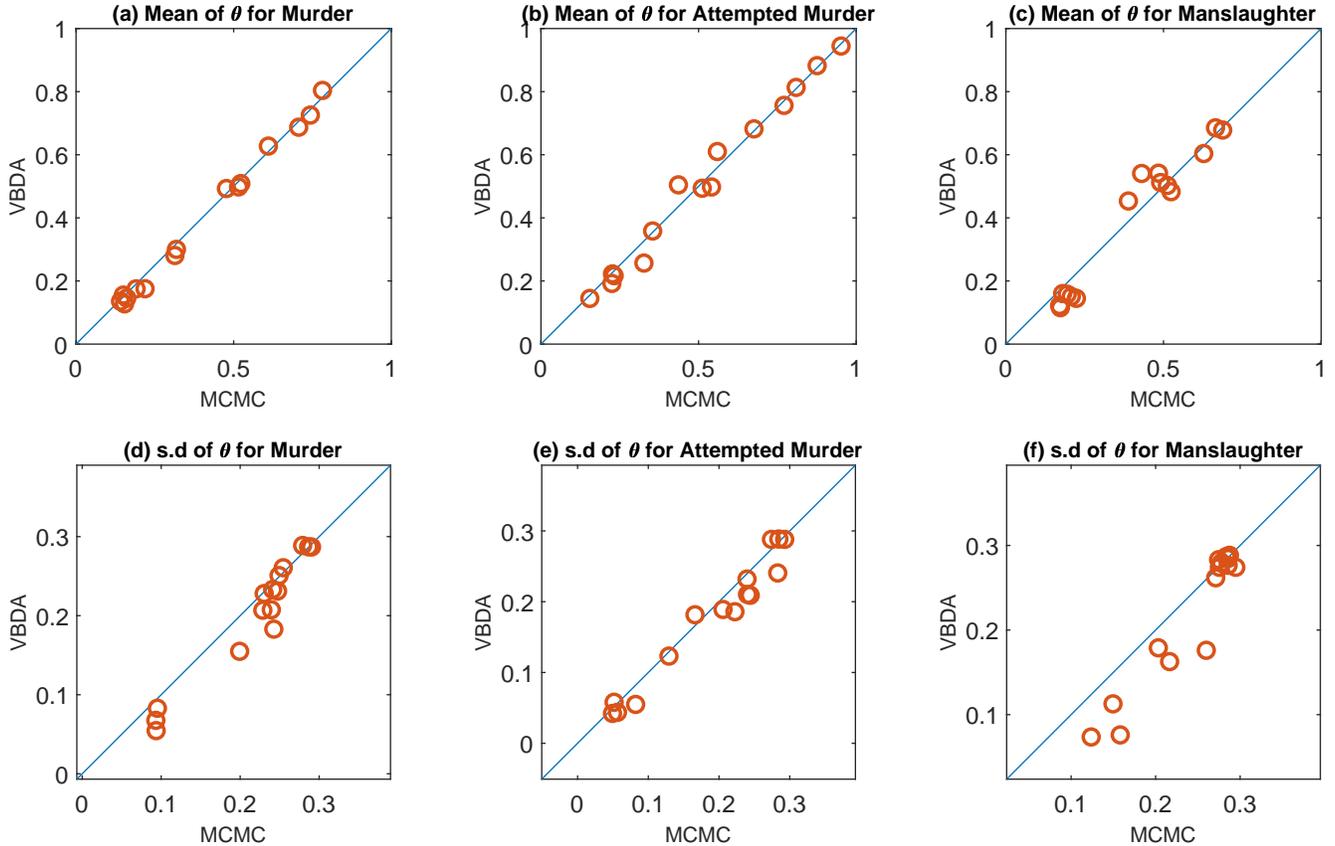}
	\end{center}
	\caption{Comparison of the parameter 
		posterior means and standard deviations from the MCMC and VBDA methods for the 
		three crime count time series.  The first row compares the posterior means, while the second row the posterior standard deviations. 
		Each column corresponds to a different crime series. VBDA was implemented with approximation VA2 and $K=3$ factors. Equivalent plots for VA1 and VA3 can be found in the Supplementary Material.}
	\label{Fig:VBvsMCMC}
\end{figure}

To illustrate the accuracy of VBDA for the three crime series, 
Figure~\ref{Fig:VBvsMCMC} plots the posterior means and standard deviations of $\bm{\theta}$ from the preferred approximation (VA2 with $K=3$)
against their (effectively exact) values computed via MCMC. 
Both moments of the VB approximations are close to those of the true posterior.
Similar plots for VA1 and VA3 (see Supplementary Materials)
suggest these are also reasonable approximations. The
first three rows in Table~\ref{tab:timetable} present the copula specifications and total estimation times for MCMC and VBDA for all examples. The computations
were undertaken on a Dell Precision workstation using Matlab, and in parallel using
8 workers for key computations for both estimators. The results show that VBDA is many times faster than
MCMC data augmentation. Moreover, the main computation of the VBDA estimator
is the repeated evaluation of  
$h$ at Step~2(b). This is slow because computing the arguments of the pair-copulas
is computationally intensive, and the
VBDA method proves even faster for 
simpler copulas.

\begin{table}
	\centering
	{\small	
		\begin{tabular}{lccccccc}
			\toprule
			Example          &  Copula   &    No. of    &   No. of   & \multicolumn{3}{c}{Time VBDA} &  Time   \\
			                 & Dimension & Pair-Copulas & Parameters &  VA1   &  VA2   &     VA3     &  MCMC   \\ \midrule
			Murder           &    264    &      3       &     15     & 0.75 h & 0.83 h &   2.30 h    & 13.85 h \\
			Attempted Murder &    264    &      3       &     15     & 0.76 h & 0.83 h &   2.32 h    & 14.51 h \\
			Manslaughter     &    264    &      3       &     15     & 0.75 h & 0.83 h &   2.32 h    & 14.66 h \\
			Auto-Logistic    &    200    &      3       &     15     & 0.62 h & 0.68 h &   1.53 h    & Failed  \\
			Homicide         &    792    &      12      &     60     & 3.89 h & 4.10 h &   16.26 h   &   ---   \\
			Bankruptcy/VIX   &    658    &      9       &     45     & 3.50 h & 3.56 h &   5.88 h    &   ---   \\ \bottomrule
		\end{tabular}
	}
	\caption{Copula model specifications and estimation times for the six examples. 
		The first four examples are univariate time series models, `Homicide' is the trivariate time
		series model in Section~\ref{sec:trivariateNSW}, and `Bankruptcy/VIX' is the bivariate time series model
		in Section~\ref{sec:vix}.
		The dimension of the D-vine copula, the number of unique pair-copulas, 
		and the number of unique copula parameters, are reported.  The total 
		times to estimate each model using our Matlab implementation (using 8 workers)
		are reported in hours. 
		The MCMC estimates are based on 30000 sweeps, 
		while the VBDA estimates are based on 5000 steps, $S = 500$ and $K=3/15$ for the 
		univariate/multivariate examples. 
		Computation times for MCMC estimation of the multivariate time series are excessive
		and unreported, while the MCMC sampler failed to converge for the Auto-Logistic example.}
	\label{tab:timetable}
\end{table} 

\vspace{-15pt}
\section{Multivariate Ordinal Time Series}\label{section:4}
\vspace{-10pt}
In this section we extend the time series copula to capture the dependence in 
multiple ordinal-valued series.

\vspace{-10pt}
\subsection{Copula Model and Estimation}
\vspace{-5pt}
Consider an $r$-dimensional stationary stochastic process $\{\bm{Y}_t\}$, where 
$\bm{Y}_t=(Y_{1,t},\ldots,Y_{r,t})'$, and each element $Y_{i,t}$ is ordinal-valued
with margin $G_i$. We assume
a time series copula model with $Tr$-dimensional copula function.
Then if $\bm{y} = \left(\bm{y}_1',\dots,\bm{y}_T'\right)'$
and 
$\bm{y}_t = \left(y_{1,t},\dots,y_{r,t}\right)'$, we estimate it using the
augmented likelihood
\begin{equation}\label{Eq:augmentedLikMulti}
f(\bm{u},\bm{y}|\bm{\theta}) = c(\bm{u}|\bm{\theta})\prod_{i=1}^{r}\prod_{t=1}^{T}{\cal I}\left(a_{i,t}\le u_{i,t}< b_{i,t}\right)\,,
\end{equation}
where $\bm{u} = \left(\bm{u}_1',\dots,\bm{u}_T'\right)'$,
$\bm{u}_t = \left(u_{1,t},\dots,u_{r,t}\right)'$, $a_{i,t}=G_i(y_{i,t}^-)$ and $b_{i,t}=G_i(y_{i,t})$.
The copula density $c$ in Equation~(\ref{Eq:augmentedLikMulti})
 captures both cross-sectional and serial dependence jointly. For this, 
\cite{biller2009} and \cite{smithvahey2015} use a Gaussian copula,
with parameter matrix equal to the correlation matrix of a stationary vector autoregression. However, a Gaussian copula cannot capture the high level of persistence in the variance often
exhibited in ordinal time series. Instead, we follow~\cite{beare15}, \cite{brechmann15}, \cite{Smith2015} and \cite{LoaizaSmithManee2017} and again use a D-vine copula, but with a parsimonious
form corresponding to a stationary Markov $p$ multivariate series. 
The pair-copula components are of the form at Equation~(\ref{eq:cmix}) to account for heteroskedasticity.

\cite{Smith2015} shows that this D-vine has a density that can be factorized as
\begin{equation}
c^{DV}(\bm{u})={\cal K}_0(\bm{u}_1)\prod_{t=2}^T\left( {\cal K}_0(\bm{u}_t) 
\prod_{k=1}^{\min(t-1,p)} {\cal K}_k(\bm{u}_{t-k},\ldots,\bm{u}_t)\right)\,.
\label{eq:multicop}
\end{equation}
The functionals ${\cal K}_0,\ldots,{\cal K}_p$ are each products of blocks of
pair-copula densities,
and do not vary with $t$ for stationary series. They are
defined as
\[
{\cal K}_k\left(\boldsymbol{u}_{t-k},\dots,\boldsymbol{u}_t\right) = 
\begin{cases}
\prod_{l_1=1}^{r}\prod_{l_2=1}^{l_1-1}c_{l_2,l_1}^{(0)}\left(u_{j|i-1},u_{i|j+1};\bm{\theta}_{l_2,l_1}^{(0)}\right)& \text{if} \ \ k=0\\
\prod_{l_1=1}^{r}\prod_{l_2=1}^{r}c_{l_2,l_1}^{(k)}\left(u_{j|i-1},u_{i|j+1};
\bm{\theta}_{l_2,l_1}^{(k)}\right)& \text{if} \ \ 1\leq k\leq p\,,
\end{cases}
\]
where $c^{(k)}_{l_2,l_1}$ is a bivariate pair-copula density 
with parameters $\bm{\theta}_{l_2,l_1}^{(k)}$. 
When $k=0$, there are $r(r-1)/2$ of these associated with ${\cal K}_0$, 
and they collectively capture
cross-sectional dependence between the $r$ variables. For example,
if they were each equal to the bivariate independence copula with 
density $c^{(0)}_{l_2,l_1}=1$, then 
${\cal K}_0=1$ and the variables would be independent
contemporaneously. When $k>p$, there
are $r^2$
pair-copulas associated with block ${\cal K}_k$ that capture serial dependence 
at lag $k$. 
In total,
there are $p(r^2)+r(r-1)/2$ unique pair-copulas, which is much less than
the $Tr(Tr-1)/2$ in an
unconstrained D-vine. 
The indices of the pair-copula 
arguments are $i=l_1+r(t-1)$ and $j=l_2+r(t-k-1)$, 
and
the argument values
$\{u_{i|j},u_{j|i};\,i=1,\ldots,Tm,\,j<i\}$ 
are computed
using the Algorithm 1 of~\cite{LoaizaSmithManee2017}.
Last, we note that if $r=1$, then ${\cal K}_0=1, i=t, j=t-k$ and 
${\cal K}_k=c_{1,1}^{(k)}(u_{t-k|t-1},u_{t|t-k+1})$, so that with the notation
$c_{k+1}\equiv c_{1,1}^{(k)}$, the copula densities at 
Equations~(\ref{eq:DvineLikeli}) and~(\ref{eq:multicop}) are the same.

To measure the dependence between $Y_{j,s}\in S_j$ and $Y_{i,t}\in S_i$, with $k=t-s$, we use the Spearman's correlation
\begin{align}\label{Eq:SpearRhoMulti}\nonumber
\rho_{i,j,k} =-3+3\sum_{y_{j,s}\in S_j}\sum_{y_{i,t}\in S_i}&g_j\left(y_{j,s}\right)g_i\left(y_{i,t}\right)\left(\bar C_{j,i,k}\left(b_{j,s},b_{i,t}\right)\right.+\bar C_{j,i,k}\left(b_{j,s},a_{i,t}\right)  +\\ 
& \left.  \bar C_{j,i,k}\left(a_{j,s},b_{i,t}\right) +\bar C_{j,i,k}\left(a_{j,s},a_{i,t}\right)\right)\,.
\end{align}
Here, $g_i$ is the probability mass function corresponding to $G_i$,
while $\bar C_{j,i,k}(u_{j,s},u_{i,t})$ is the copula function
of the bivariate marginal of $(Y_{j,s},Y_{i,t})$. The latter is computed by simulating from
 $c^{DV}$ and then constructing the empirical copula function for 
$\bar C_{j,i,k}$.

The augmented posterior of this copula time series model is
\begin{equation}\label{Eq:posteriorAugMulti}
p(\bm{u},\bm{\theta}|\bm{y}) \propto c^{DV}(\bm{u}|\bm{\theta})p\left(\bm\theta\right)\prod_{i=1}^{r}\prod_{t=1}^{T}{\cal I}\left(a_{i,t}\le u_{i,t}< b_{i,t}\right)\,.
\end{equation}
Because of the very large number of elements in $\bm u$, 
estimation using MCMC is computationally infeasible for  
even moderate values of $r$ and $T$.
However, our VBDA estimator can be employed with the same variational approximations
outlined in Section~\ref{sec:va}. We note that in our empirical work we employ VA3 as exactly outlined, although
the sparse pattern of
$\Omega^{-1}$ can be further tailored to match the possible dependence
structure of $p(\bm{u}|\bm{y})$ for this case.

\begin{table}[H]
	\centering
	{\small
		\begin{tabular}{ccccccc}
			\toprule
			&              \multicolumn{6}{c}{D-Vine copula: $C^{MIX}$ with Convex Gumbel Components}               \\
			Parameters         & $\tau^a>0$ & $\delta^a$ & $\tau^b>0$ & $\delta^b$ &  $w$  &                 Spearman                  \\ \midrule
			$\bm{\theta}_{1,2}^{(0)}$ &   0.200    &   0.548    &   0.347    &   0.507    & 0.929 & 0.169      {\footnotesize( 0.109,0.230)}  \\
			$\bm{\theta}_{1,3}^{(0)}$ &   0.084    &   0.395    &   0.118    &   0.499    & 0.617 & 0.020     {\footnotesize( -0.026,0.079)}  \\
			$\bm{\theta}_{2,3}^{(0)}$ &   0.153    &   0.656    &   0.167    &   0.496    & 0.785 & 0.093     {\footnotesize(  0.035,0.159)}  \\ \midrule
			$\bm{\theta}_{1,1}^{(1)}$ &   0.106    &   0.356    &   0.157    &   0.475    & 0.586 & 0.009     {\footnotesize( -0.055,0.073)}  \\
			$\bm{\theta}_{1,2}^{(1)}$ &   0.173    &   0.476    &   0.118    &   0.488    & 0.666 & 0.076     {\footnotesize(  0.014,0.154)}  \\
			$\bm{\theta}_{1,3}^{(1)}$ &   0.104    &   0.414    &   0.133    &   0.505    & 0.505 & -0.006     {\footnotesize( -0.070,0.055)} \\
			$\bm{\theta}_{2,1}^{(1)}$ &   0.205    &   0.706    &   0.167    &   0.482    & 0.842 & 0.154     {\footnotesize(  0.091,0.223)}  \\
			$\bm{\theta}_{2,2}^{(1)}$ &   0.341    &   0.623    &   0.220    &   0.474    & 0.932 & 0.306     {\footnotesize(  0.240,0.373)}  \\
			$\bm{\theta}_{2,3}^{(1)}$ &   0.153    &   0.499    &   0.110    &   0.528    & 0.528 & 0.023     {\footnotesize( -0.037,0.095)}  \\
			$\bm{\theta}_{3,1}^{(1)}$ &   0.130    &   0.476    &   0.097    &   0.477    & 0.512 & 0.012     {\footnotesize( -0.040,0.075)}  \\
			$\bm{\theta}_{3,2}^{(1)}$ &   0.172    &   0.417    &   0.143    &   0.485    & 0.698 & 0.084     {\footnotesize(  0.022,0.156)}  \\
			$\bm{\theta}_{3,3}^{(1)}$ &   0.150    &   0.555    &   0.136    &   0.498    & 0.568 & 0.028     {\footnotesize( -0.034,0.104)}  \\ \bottomrule
		\end{tabular}
	}
	\caption{The VBDA posterior means of the pair-copula parameters
		for the D-Vine copula fitted to the three-dimensional crime series using approximation VA2 and $K=15$ factors. 
		The posterior mean and 90\% probability intervals
		are also given for the
		Spearman's rho of each pair copula.
		Murder, Attempted Murder and Manslaughter counts
		are denoted
		as series 1, 2 and 3, respectively. Estimates for VA1 and VA3 are similar, and are given in the
		Supplementary Material.}
	\label{Tab:CopulaEstimatesCrimes}
\end{table}

\vspace{-10pt}
\subsection{Example: New South Wales Homicide}\label{sec:trivariateNSW}
\begin{figure}
	\begin{center}
		\includegraphics[scale =0.6]{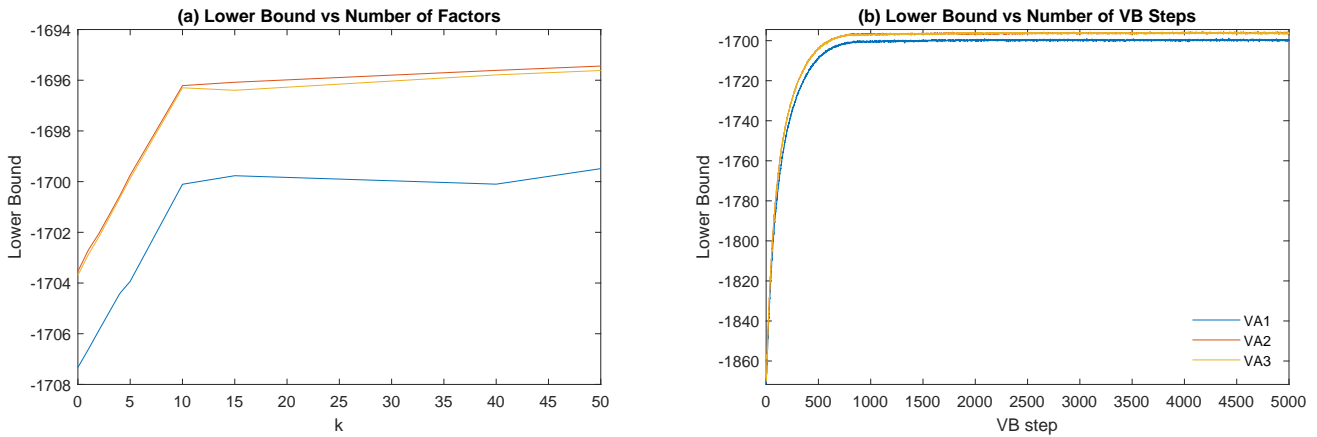}
	\end{center}
	\caption{Variational lower bound values ${\cal L}(\bm{\lambda})$ for the VBDA estimators of the  trivariate time series Homicide example. Panel~(a) plots ${\cal L}$ at 
		the estimate of $\bm{\lambda}$, against the number of factors $K$ in the
		factor decomposition of $\Sigma$. 
		Panel~(b) plots ${\cal L}$ against VB step for the case of $K=15$ factors. In both panels 
		results are given for the
		VA1 (blue line), VA2 (red line) and VA3 (orange line) approximations.}
	\label{Fig:LowerBoundHomicideMultivariate}
\end{figure}

\vspace{-5pt}
We consider a trivariate time series copula model for the NSW monthly crime counts, with 
the empirical distributions as univariate marginals.  
The copula density is given in
Equation~(\ref{eq:multicop}), where we set $p=1$ and adopt 
pair-copula densities of the form $c^{MIX}$. 
The dimension of the D-vine copula is $3 \times 264 = 792$, 
and Table~\ref{tab:timetable} reports its specification. The copula parameters are 
estimated using the VB estimator with
$K = 0,1,2,4,5,10,15,40,50$ factors.  
To estimate the gradient $S=500$ in Algorithm~1, and 5000 VB steps are used with $K=15$.
Figure~\ref{Fig:LowerBoundHomicideMultivariate} plots the variational lower bound 
against $K$ in panel~(a), and against the VB step when $K=15$ in panel~(b), for VA1, VA2 and VA3.
 A total of 
$K=15$ factors appears sufficient, while both VA2 and VA3 give similar results, but dominate VA1. 
Table~\ref{Tab:CopulaEstimatesCrimes} reports the posterior means 
and intervals of $\bm{\theta}$ for VA2, although those for VA1 and VA3 are very similar and are reported
in the Supplementary Material.

\begin{table}
	\centering
	\scalebox{1}{
		\begin{tabular}{lcccc}
			\toprule
			&          Murder$_t$          &    Attempted Murder$_t$     &       Manslaughter$_t$       &  \\ \midrule
			\multicolumn{4}{l}{$\underline{k=0}$}                                                                                &  \\
			Attempted Murder$_t$     &           0.246***           &              -              &              -               &  \\
			& ({\footnotesize 0.16,0.33})  &                             &                              &  \\
			Manslaughter$_t$         &            0.055*            &          0.119***           &              -               &  \\
			& ({\footnotesize-0.01,0.13})  &  ({\footnotesize0.05,0.2})  &                              &  \\
			\multicolumn{4}{l}{$\underline{k=1}$}                                                                                &  \\
			Murder$_{t-1}$           &            0.071*            &          0.219***           &            0.028             &  \\
			& ({\footnotesize-0.02,0.17})  & ({\footnotesize0.12,0.33})  & ({\footnotesize-0.05,0.10})  &  \\
			Attempted Murder$_{t-1}$ &           0.221***           &          0.470***           &           0.089**            &  \\
			& ({\footnotesize 0.13,0.32})  & ({\footnotesize0.39,0.55})  & ({\footnotesize 0.01,0.18})  &  \\
			Manslaughter$_{t-1}$     &            0.015             &           0.104**           &            0.046             &  \\
			& ({\footnotesize -0.05,0.09}) & ({\footnotesize 0.03,0.19}) & ({\footnotesize -0.02,0.13}) &  \\ \bottomrule
		\end{tabular}}
		\caption{The VBDA estimates of the Spearman pairwise correlations $\rho_{i,j,k}$ for $k = 0,1$, using approximation VA2. The estimates of the posterior means are reported, with the 90\% posterior intervals below. These are computed from the copula model by simulation. 
			Correlations with approximate posterior intervals that do not contain zero
			at the 10\%, 5\% and 1\% level are denoted with `*', `**' and `***', respectively.}
		\label{tab:spearmansCorrGumbelMulti}
	\end{table}      

\begin{figure}
	\begin{center}
		\includegraphics[scale =0.55]{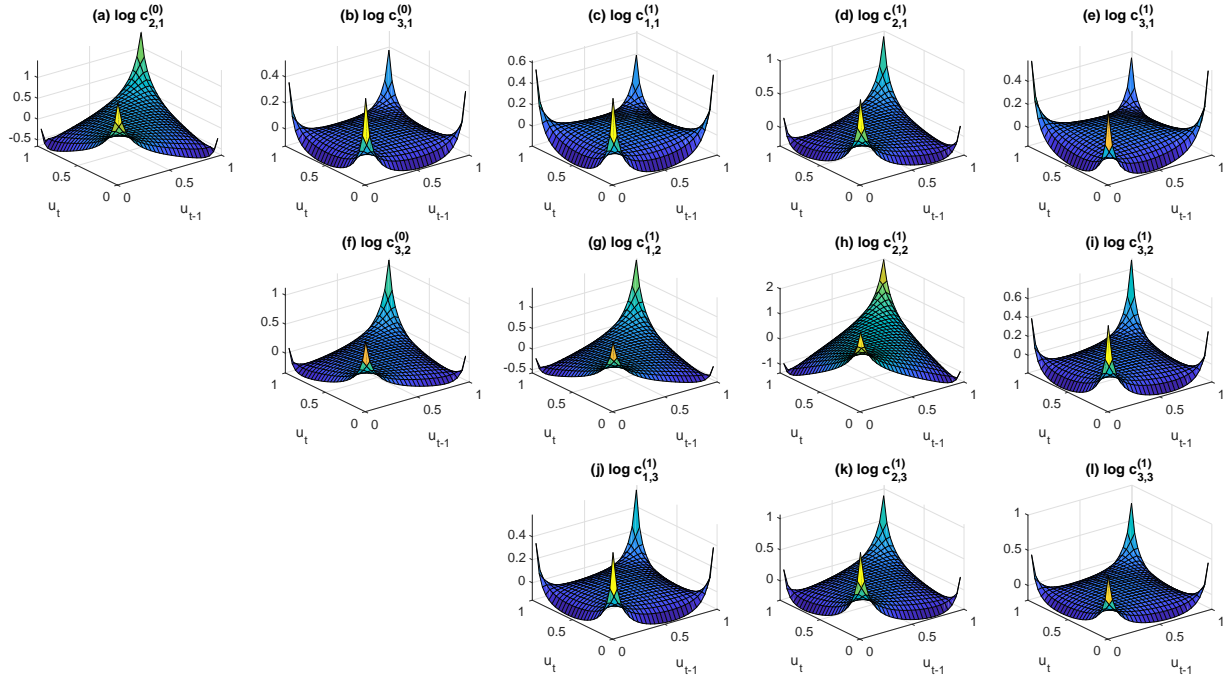}
	\end{center}
	\caption{Logarithm of the pair-copula densities computed at the VBDA posterior mean parameter values for the trivariate Homicide example. 
		For each density, the superscript indicates the lag in the second argument of the pair-copula, while the subscript labels 1, 2 and 3 correspond to the Murder, Attempted Murder and Manslaughter count series, respectively. Results are given for variational approximation VA2 with $K=3$ factors.}
	\label{Fig:PairCopulasCrimeMulti}
\end{figure}

Table~\ref{tab:spearmansCorrGumbelMulti} reports the
estimates of the pairwise Spearman correlations. The 
contemporaneous correlations ($k=0$) are given in the top two rows, and first order serial correlations
($k=1$) in the bottom rows.
There is positive 
contemporaneous correlation between Attempted Murder and Murder, and also (weakly) with Manslaughter. There is first order serial correlation in
Attempted Murder, but not in the other two crimes. The most striking result is that 
Attempted Murder is positively correlated with Murder and Manslaughter one month later, suggesting
it is a leading indicator of these two crimes.
However, these correlations measure dependence in the level only.  
Figure~\ref{Fig:PairCopulasCrimeMulti} displays the logarithm of
the 15 unique pair-copula densities. 
Most have mass in the off-diagonal corners of the unit square, indicating
that the copula is capturing heteroskedasticity and `variance spill-overs'
between the three series. The 3 pair-copulas on the lefthand side capture
contemporaneous cross-sectional dependence. The 9 pair-copulas on the righthand side 
capture first order serial dependence. For example, $c_{2,1}^{(1)}$ is very far from
uniform, and captures strong variance spill-over between Attempted Murder and Murder.

\vspace{-15pt}
\section{Mixed Multivariate Time Series}\label{section:5}
\vspace{-10pt}
\subsection{Copula Model and Estimation}
\vspace{-5pt}
Consider the case of a stochastic process $\{\bm{Y}_t\}$, where
$\bm{Y}_t=(Y_{1,t},\ldots,Y_{r,t})'$ consists
of $d$ ordinal and $r-d$ continuous-valued variables, which we refer to as `mixed'. A copula model using the same D-vine with density 
$c^{DV}$ at Equation~(\ref{eq:multicop}) can be used 
to construct a flexible time series model. 
Without loss of generality,
if the first $d$ elements of $\bm{Y}_t$ are ordinal,
then \cite{SmithKhaled2012} suggest estimation of a copula model with mixed margins using the 
augmented density at Equation~(\ref{Eq:augmentedLikMulti}), but
where $f(y_{i,t}|u_{i,t})={\cal I}(u_{i,t}=G_i(y_{i,t}))$ is a point
mass for $i=d+1,\ldots,r$. They
discuss how to implement MCMC data augmentation,
but this approach can be slow or computationally
infeasible for values of $Td$ that occur frequently in time series analysis.

Let ${\cal C}=\{(i,t): i=d+1,\ldots,r\,; t=1,\ldots,T\}$ denote
the indicies of the continuous-valued $Y_{i,t}$, and 
$\bm u_{D}$ be the $Td$ latents corresponding
to the ordinal variables. Then 
VBDA can be employed using the variational
approximations outlined in Section~\ref{sec:va}, but where
$q_{\lambda^b}(\bm{u})=
\tilde q_{\lambda^b}(\bm{u}_D)\prod_{(i,t)\in {\cal C}}{\cal I}(u_{i,t}=G_i(y_{i,t}))$, and 
approximations VA1 to VA3 are considered for $\tilde q_{\lambda^b}$. 
Algorithm~\ref{alg:VB} can be used to approximate
the augmented posterior, but where $u_{i,t}=G_i(y_{i,t})$ are constants
for $(i,t)\in {\cal C}$, and are not generated.

Equation~(\ref{Eq:SpearRhoMulti}) can be used to compute the Spearman correlation $\rho_{i,j,k}$  between two 
ordinal-valued variables $(Y_{j,s},Y_{i,t})$ with $k=t-s$ and $s<t$. If  both variables are 
continuous-valued, then
$\rho_{i,j,k}=12\int \bar C_{i,j,k}(u,v)\mbox{d}u\mbox{d}v-3$. But if $Y_{j,s}\in S_j$ is ordinal and $Y_{i,t}$ is
continuous, then 
\[
\rho_{i,j,k}=
6\sum_{y_{j,s}\in S_j}g_j\left(y_{j,s}\right)\int g_i\left(y_{i,t}\right)\left(\bar C_{j,i,k}\left(b_{j,s},G_i(y_{i,t})\right) + \bar C_{j,i,k}\left(a_{j,s},G_i(y_{i,t})\right)\right)\mbox{d}y_{i,t} -3\,,
\]
where the integral can be computed numerically. In all cases, $\bar C_{i,j,k}$
is evaluated by simulation as previously.

\begin{figure}
	\begin{center}
		\includegraphics[scale =0.65]{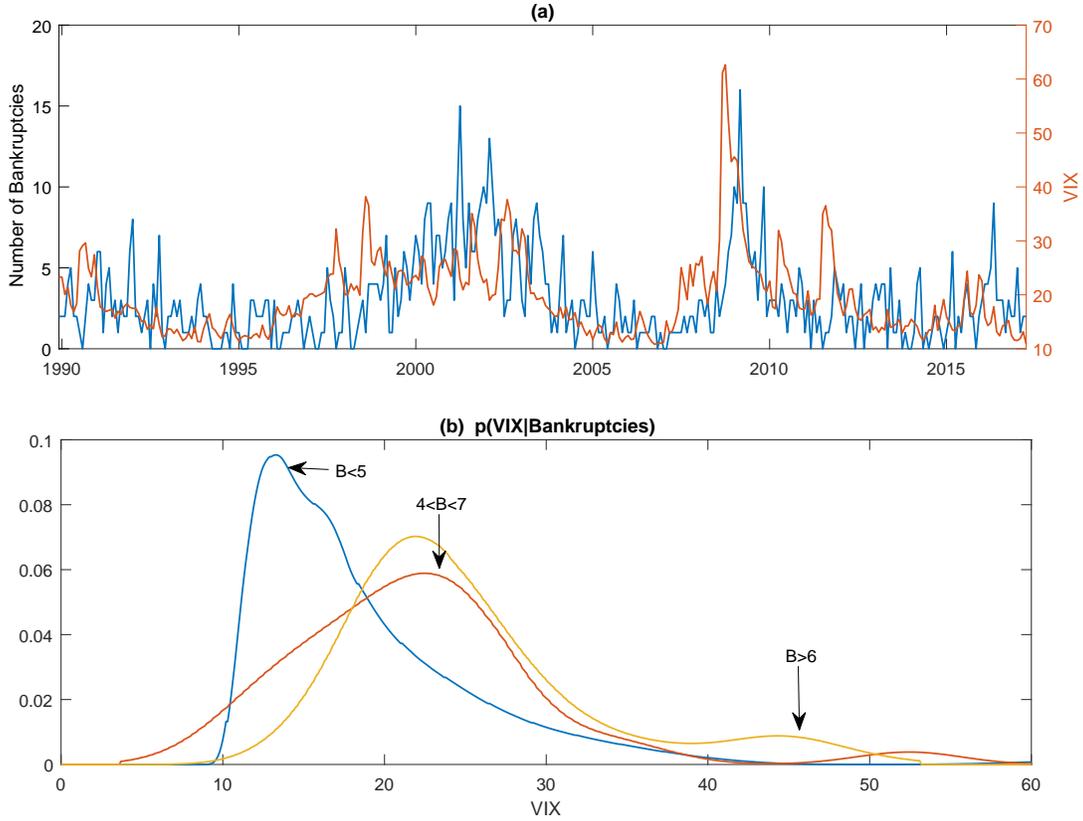}
	\end{center}
	\caption{Panel (a) plots the bankruptcy count and VIX time series. The left vertical
		axis displays the number of bankruptcies, while the right vertical axis displays the VIX values. 
		Panel~(b) plots the density estimates of VIX, conditional on three ranges of values for 
		the number of bankruptcies. Denoting $B$ as the number of bankruptcies, the ranges
		are $B<5$ (blue line density), $4<B<7$ (red line density) and $B>6$ (yellow line density).}
	\label{Fig:BankruptcySeries}
\end{figure}

\begin{table}
	\centering
	\scalebox{1}{
		\begin{tabular}{lcc}
			\toprule
			&       Bankruptcy$_t$       &          VIX$_t$           \\ \midrule
			VIX$_t$            &          0.102***           &                            \\
			& ({\footnotesize 0.04,0.16}) &                            \\
			Bankruptcy$_{t-1}$ &          0.448***          &           0.075*           \\
			& ({\footnotesize0.37,0.52}) & ({\footnotesize-0.01,0.16}) \\
			VIX$_{t-1}$        &          0.200***          &          0.862***          \\
			& ({\footnotesize0.14,0.27}) & ({\footnotesize0.83,0.89}) \\
			Bankruptcy$_{t-2}$ &          0.334***          &           0.079*           \\
			& ({\footnotesize0.24,0.43}) & ({\footnotesize-0.02,0.18}) \\
			VIX$_{t-2}$        &          0.238***          &          0.740***          \\
			& ({\footnotesize0.16,0.31}) & ({\footnotesize0.67,0.80}) \\ \bottomrule
		\end{tabular}}
		\caption{The VB posterior means of the Spearman unconditional pairwise correlations $\rho_{i,j,k}$,
			for $k = 0,1,2$, computed from the copula model via simulation. Correlations with (variational) posterior intervals that do not contain zero
			at the 10\%, 5\% and 1\% level are denoted with `*', `**' and `***', respectively.}
		\label{tab:spearmansCorrGumbelMultiBankrutcy}
	\end{table}          

\begin{figure}
	\begin{center}
		\includegraphics[scale =0.5]{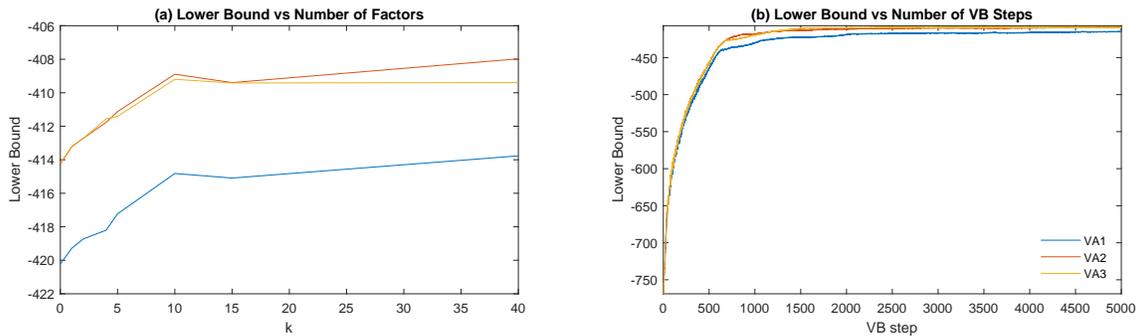}
	\end{center}
	\caption{Variational lower bound values ${\cal L}(\bm{\lambda})$ for the VBDA estimators of the 
		bivariate time series of bankruptcy counts and VIX. Panel~(a) plots ${\cal L}$ at 
		the estimate of $\bm{\lambda}$, against the number of factors $K$ in the
		factor decomposition of $\Sigma$. 
		Panel~(b) plots ${\cal L}$ against VB step for the case of $K=15$ factors. In both panels 
		results are given for the
		VA1 (blue line), VA2 (red line) and VA3 (orange line) approximations.}
	\label{Fig:LowerBoundBankVIX}
\end{figure}

\vspace{-10pt}
\subsection{Example: Bankruptcy and the VIX}\label{sec:vix}
\vspace{-5pt}
We study the dependence between the continuous-valued
VIX index, which measures
U.S. market volatility, and the number of public company bankruptcy cases filed
in U.S. courts.
We employ the monthly average value of the VIX
obtained from the FRED website, while the bankruptcies are monthly counts
sourced from the UCLA-LoPucki Bankruptcy Research Database. The time series
are 
from
December 1989 to April 2017, so that $T=329$.
A positive relationship between market volatility and bankruptcies has been documented previously   \citep{bauer2014hazard}. Figure~\ref{Fig:BankruptcySeries}(a) plots both series, 
while Figure~\ref{Fig:BankruptcySeries}(b) displays the empirical
distribution of the VIX, conditional on the number of bankruptcies.
The positive correlation between the two series
is apparent in both panels.

We employ the D-Vine copula model with $p=2$ and a total number of $5\times9 = 45$ parameters;
Table~\ref{tab:timetable} reports the copula
specification. Separate 
variational approximations with $K = 0,1,2,4,5,10,15,40,45$ factor decompositions
for $\Sigma$ were estimated. The 
same values for $S$ and number of VB steps were adopted as in Section~\ref{sec:trivariateNSW}.
Figure~\ref{Fig:LowerBoundBankVIX}(a) plots the lower bound against $K$, and it varies
little for $K\geq 15$. This is consistent with the empirical results in~\cite{ong2017gaussian}, who
found that a higher number of factors are needed for the accurate approximation of more complex posteriors.

The copula parameter estimates are reported in the Online Appendix, while
Table~\ref{tab:spearmansCorrGumbelMultiBankrutcy} reports the 
pairwise Spearman correlations. Both the number of bankruptcies and the VIX exhibit
serial correlation, although the latter more so. The two series are positively correlated, both 
contemporaneously and in the lagged values.
However, the lagged values of the VIX are more highly correlated with later bankruptcies,
suggesting that the VIX is a leading indicator of public company bankruptcy filings. 
The fitted pair-copulas densities are plotted
in the Online Appendix, and their form
is consistent with heteroskedastic time series. For example, $c^{(1)}_{2,1}$ and 
$c^{(1)}_{1,2}$ have mass concentrated in all four corners, indicating
positive cross-correlation in the variance of the two series at different lags;
ie. volatility `spillover'. 

\vspace{-15pt}
\section{Discussion}\label{section:6}
\vspace{-10pt}
This paper makes two main contributions. The first is to propose a new
VB estimator for copula models with discrete, or a combination of discrete
and continuous, variables. The approach can be used to estimate
copulas with a higher dimension and number of parameters than previous methods. We illustrate this using time series copulas of
up to 792 dimensions and 60 parameters, although the method can be used to estimate
copula models for cross-sectional, longitudinal or spatial data 
just as readily.
The second main contribution of the paper is
to propose a new time series model for multivariate ordinal-valued variables, where
a copula captures serial and cross-sectional dependence jointly. 
Our proposed
copula is a parsimonious
D-vine that can capture serial dependence in both the level and 
conditional variance, with the latter being an important feature
in much ordinal data. The time series
model is highly flexible, where any marginal features in a time
series can
be captured by an arbitrary distribution, and is easily extended to a combination of 
discrete and continuous-valued series.
 
Ordinal time series frequently
exhibit both serial dependence in the level and heteroskedasticity.
Few existing copulas can capture both jointly, yet the D-vine used here
can do so for Markov and stationary series. 
An advantage of such copula
time series models is that they allow for the more accurate modeling of data with 
multi-modal and irregular margins, 
as well as being readily extended to multivariate series. However,
their estimation is computationally challenging using previous methods, 
and our VBDA estimator provides a new and effective solution, as illustrated by our
examples.

In our VB approach, a key observation is that it is computationally advantageous to employ
a variational approximation to the augmented posterior $p(\bm{u},\bm{\theta}|\bm{y})$,
rather than the intractable posterior $p(\bm{\theta}|\bm{y})$. This is consistent with
\cite{tan2017}, \cite{hui2017} and \cite{ong2017gaussian}, who also find that 
variational approximations to the posterior of latent variables can also 
be computationally efficient in mixed effects 
generalized linear models.
The empirical examples illustrate that the
 approximations proposed here provide a balance between computational
efficiency and accuracy.

Last, we outline some promising directions for future research.
First, \cite{gunawan2018} give an augmented likelihood
for copula models where the margins have mixed densities (not to be confused
with a combination of continuous and discrete variables). 
Extending our VBDA approach to this case would provide a
faster estimator than MCMC.
Second, copula models
for discrete spatial data are growing in popularity
\citep{hughes2015,deoliveira2018}. However, estimation is challenging 
for a large number of spatial locations, and
VBDA provides a solution. When employing 
VA3, $\Omega$ can be tailored to each case; for example,  
a natural choice for $\Omega^{-1}$ for data located 
on a regular lattice is the precision matrix of a Gaussian Markov random field.
Last,  copulas constructed by inversion of existing distributions
are popular, including those that have intractable copula 
functions $C$ and densities $c$; see \cite{smith2012,SmithMan2016} and~\cite{oh2017} for examples. Extending our VBDA estimator to such intractable copula
models for 
discrete data is an interesting extension.
\appendix
\vspace{-15pt}
\section{ADADELTA}\label{Append:Adadelta}
\vspace{-10pt}
The learning rate $\rho^{(k)}$ can be set using different methods. For example \cite{tran2017variational} set it as a sequence with manually tuned parameters. \cite{ong2016variational} propose an adaptive learning rate based on previous work by \cite{ranganath2013adaptive}. Here, we employ the ADADELTA method of \cite{zeiler2012adadelta},  which provides reliable convergence of the SGA algorithm. 
This method consists of individually updating the step size for each element in $\bm{\lambda}$ as
$$\lambda_i^{(k+1)} = \lambda_i^{(k)}+\Delta\lambda_i^{(k)}$$
with $\Delta\lambda_i^{(k)} = \rho_i^{(k)}g_{\lambda_i}^{(k)}$, $g_{\lambda_i}^{(k)}$ denoting the $i^{th}$ element of $\widehat{\nabla_{\bm{\lambda}}\mathcal{L}\left(\bm{\lambda}^{(k)}\right)}$ and $\rho_i^{(k)}$ is given by
$$\rho_i^{(k)}=\frac{\sqrt{E\left(\Delta_{\lambda i}^2\right)^{(k-1)}+\epsilon}}{\sqrt{E\left(g_{\lambda_i}^2\right)^{(k)}+\epsilon}}$$
where $\epsilon$ is a small scalar and $E\left(\Delta_{\lambda i}^2\right)^{(k)}$ and $E\left(g_{\lambda_i}^2\right)^{(k)}$ are recursively updated as
\begin{align*}
E\left(\Delta_{\lambda i}^2\right)^{(k)}=&\zeta E\left(\Delta_{\lambda i}^2\right)^{(k-1)}  +(1-\zeta)\Delta\lambda_i^{(k)^2}\\
E\left(g_{\lambda_i}^2\right)^{(k)}=&\zeta E\left(g_{\lambda_i}^2\right)^{(k-1)} +(1-\zeta)g_{\lambda_i}^{(k)^2}
\end{align*}
For the VB applications here, \cite{ong2017gaussian} is followed, and we set $\epsilon =10^{-6}$, $\zeta=0.95$, $E\left(\Delta_{\lambda i}^2\right)^{(0)}=0$ and $E\left(g_{\lambda_i}^2\right)^{(0)}=0$.

\vspace{-15pt}
\section{Proof of Theorem~1}\label{append:proof}
\vspace{-10pt}
The expression at (a) is obtained simply by integrating over Equation~(\ref{Eq:postaugmented})
with respect to $\bm{\theta}$. To derive the expression at~(b), first note that 
$p(\bm{u}|\bm{\theta},\bm{y})\propto \prod_{t=1}^T {\cal I}(a_t\leq u_t < b_t)c(\bm{u}|\bm{\theta})$, so that
\begin{eqnarray*}
p(u_t|\bm{\theta},\bm{y}) &=& \int p(\bm{u}|\bm{\theta},\bm{y})\mbox{d}\bm{u}_{s\neq t} \\
 & \propto & {\cal I}(a_t\leq u_t < b_t) \int c(\bm{u}|\bm{\theta}) \prod_{s\neq t} {\cal I}(a_s\leq u_s < b_s)\mbox{d}\bm{u}_{s\neq t}\\
 &= &{\cal I}(a_t\leq u_t < b_t) A(u_t|\bm{\theta})\,,
\end{eqnarray*}
where $A(u_t|\bm{\theta})$ is as defined in Theorem~1.
Therefore, 
\[
p(u_t|\bm{y})=\int p(u_t|\bm{\theta},\bm{y})p(\bm{\theta}|\bm{y})\mbox{d}\bm{\theta}\propto {\cal I}(a_t\leq u_t < b_t) 
\int A(u_t|\bm{\theta})p(\bm{\theta}|\bm{y})\mbox{d}\bm{\theta}\,,
\]
which is intractable.

\vspace{-15pt}
\section{Derivatives}\label{append:nabla}
\vspace{-10pt}
In this appendix we compute the gradient
$\nabla_{\bm{\lambda}}\log q_\lambda(\bm{\theta},\bm{u})$ to implement Steps~1(b) and~2(b) for Algorithm~\ref{alg:VB}.
To  present these succinctly the following notation is introduced. For a matrix $A$ of dimension $n\times K$, the function $\text{vech}(.)$ is defined as $\text{vech}(A)= \left(A_{1:n,1}',\dots,A_{K:n,K}'\right)'$ with $A_{k:n,k} = \left(A_{k,k},\dots,A_{n,k}\right)'$ for $k = 1,\dots,K$. Also, the vector of diagonal entries of the square matrix $Z$ is written as $\text{diag}(Z)$. 
Employing this notation, the vector of parameters $\bm{\lambda}^a$ can be written as $\bm{\lambda}^a = \left(\bm\mu',\bm{b}',\bm{d}'\right)'$ with $\bm b = \text{vech}(B)$, and the gradient 
$\nabla_{\lambda^a} \text{log}\left(q_{\lambda^a}\left(\bm{\theta}\right)\right)=
\left(\nabla_{\mu}\text{log}\left(q_{\lambda^a}\left(\bm{\theta}\right)\right)',
\nabla_{b}\text{log}\left(q_{\lambda^a}\left(\bm{\theta}\right)\right)',
\nabla_{d}\text{log}\left(q_{\lambda^a}\left(\bm{\theta}\right)\right)'\right)'$
where
\begin{align*}
\nabla_{\mu}\text{log}\left(q_{\lambda^a}\left(\bm{\theta}\right)\right) =& \left(B'B+D\right)^{-1}\left(\bm{\theta}-\bm{\mu}\right)                    \\
\nabla_{b}\text{log}\left(q_{\lambda^a}\left(\bm{\theta}\right)\right) =&  \text{vech}\left(-\left(B'B+D^2\right)^{-1}B+\left(B'B+D^2\right)^{-1}\left(\bm{\theta}-\bm{\mu}\right)\left(\bm{\theta}-\bm{\mu}\right)'\left(B'B+D^2\right)^{-1}B\right)                   \\
\nabla_{d}\text{log}\left(q_{\lambda^a}\left(\bm{\theta}\right)\right) =&\text{diag}\left(-\left(B'B+D^2\right)^{-1}D+\left(B'B+D^2\right)^{-1}\left(\bm{\theta}-\bm{\mu}\right)\left(\bm{\theta}-\bm{\mu}\right)'\left(B'B+D^2\right)^{-1}D\right)\,. 
\end{align*}
Fast calculation of these gradients can be undertaken using the Woodbury formula; see~\cite{ong2017gaussian} for further details.

For VA1, $\lambda^b=\emptyset$, so that
$\nabla_{\lambda^b} \log q_{\lambda^b}\left(\bm{u}\right)=0$. 
For VA2, if $c_t=\log \omega_t$, then 
\[
\log q_{\lambda^b}(\bm{u})=\sum_{t=1}^T\left(\frac{1}{2}z_t^2 -c_t
 -\frac{(z_t-\eta_t)^2}{2\exp(2c_t)}-\log(b_t-a_t) \right)\,,
\]
with derivatives $\nabla_{c_t} \log q_{\lambda^b}\left(\bm{u}\right)=
\exp(-2c_t)(z_t-\eta_t)^2-1$ and 
$\nabla_{\eta_t} \log q_{\lambda^b}\left(\bm{u}\right)=(z_t-\eta_t)/\omega^2_t$. Last,
for VA3, if $\Omega^{-1}=LL'$, then
\[
\log q_{\lambda^b}(\bm{u})=\log |L| -\frac{1}{2}(\bm{z}-\bm{\eta})'LL'(\bm{z}-\bm{\eta})+\sum_{t=1}^T \frac{1}{2}z_t^2-\log(b_t-a_t)\,,
\]
with derivatives 
$\nabla_{\bm{\eta}} \log q_{\lambda^b}\left(\bm{u}\right)=(\bm{z}-\bm{\eta})'LL'$, and 
\[
\nabla_{\mbox{\footnotesize vec}(L)} \log q_{\lambda^b}\left(\bm{u}\right)=
\mbox{vec}((L^{-1})')'-\frac{1}{2}\left( (\bm{z}-\bm{\eta})'\otimes (\bm{z}-\bm{\eta})\right)(I_{T^2}+K_{T,T})
(L\otimes I_T)
\]
where $K_{T,T}$ is a commutation matrix. Note that the gradient is for a full factor $L$, although
for the sparse $L$ employed here we compute $\nabla_{\mbox{\footnotesize vec}(L)} \log q_{\lambda^b}\left(\bm{u}\right) $
 using sparse matrix operations in Matlab, and only evaluate it
for the non-zero elements of $L$. Last, to derive this derivative we have used the identity $\frac{\partial}{\partial \mbox{\footnotesize vec}(A)}|A|=\mbox{vec}(|A|(A^{-1})')'$ for invertible square matrix $A$.

\baselineskip14pt 
\bibliography{Project_bib}
\bibliographystyle{apa}
\end{document}